\documentclass[useAMS,usenatbib]{mn2e}

\usepackage{psfig}
\usepackage{graphicx}
\usepackage{times}
\usepackage{natbib}

\newif\ifAMStwofonts
\AMStwofontstrue

\newcommand{\be}{\begin{equation}}
\newcommand{\ee}{\end{equation}}
\newcommand{\ba}{\begin{eqnarray}}
\newcommand{\ea}{\end{eqnarray}}
\newcommand{\brr}{\begin{array}}
\newcommand{\err}{\end{array}}
\newcommand{\bc}{\begin{center}}
\newcommand{\ec}{\end{center}}
\newcommand{\hm}{\,h^{-1}{\rm Mpc}}

\newcommand{\msun}{\,h^{-1}{\rm M}_\odot}

\newcommand{\vel}{\,{\rm km\,s^{-1}}}

\newcommand{\mincir}{\raise
  -2.truept\hbox{\rlap{\hbox{$\sim$}}\raise5.truept \hbox{$<$}\ }}
\newcommand{\magcir}{\raise
  -2.truept\hbox{\rlap{\hbox{$\sim$}}\raise5.truept \hbox{$>$}\ }}
\newcommand{\siml}{\raise
  -2.truept\hbox{\rlap{\hbox{$\sim$}}\raise5.truept \hbox{$<$}\ }}
\newcommand{\simg}{\raise
  -2.truept\hbox{\rlap{\hbox{$\sim$}}\raise5.truept \hbox{$>$}\ }}

\title[Galaxy population in simulations of clusters]
{Properties of the galaxy population in hydrodynamical
  simulations of clusters}
\author[A. Saro, et al.]
{A. Saro$^{1,2}$, S. Borgani$^{1,2,3}$, L. Tornatore$^{4,2}$,
 K. Dolag$^5$, G. Murante$^{6,1}$,
\\ ~\\
\LARGE{\rm A. Biviano$^3$, F. Calura$^3$ \& S. Charlot$^7$
}\\ ~\\
$^1$ Dipartimento di Astronomia dell'Universit\`a di Trieste, via
  Tiepolo 11, I-34131 Trieste, Italy (saro,borgani@oats.inaf.it)\\
$^2$ INFN -- National Institute for Nuclear Physics, Trieste,
  Italy\\ 
$^3$ INAF, Osservatorio Astronomico di Trieste, via Tiepolo 11,
  I-34131 Trieste, Italy (calura,biviano@oats.inaf.it)\\
$^4$ SISSA/International School for Advanced Studies, via Beirut 4,
 34100 Trieste, Italy (torna@sissa.it)\\
$^5$ Max-Planck-Institut f\"ur Astrophysik, Karl-Schwarzschild Strasse
  1, Garching bei M\"unchen, Germany (kdolag,charlot@mpa-garching.mpg.de)\\
$^6$ INAF, Osservatorio Astronomico di Torino, Strada Osservatorio 20,
  I-10025 Pino Torinese, Italy (giuseppe@to.astro.it)\\
$^7$ Institut d'Astrophysique de Paris, CNRS, 98 bis Boulevard Arago,
  Paris 75014, France (charlot@iap.fr)}
\begin{document}

\date{Accepted ???. Received ???; in original form ???}

\maketitle

\label{firstpage}

\begin{abstract}
  We present a study of the galaxy population predicted by
  hydrodynamical simulations of galaxy clusters. These simulations,
  which are based on the {\small GADGET-2} Tree+SPH code, include gas
  cooling, star formation, a detailed treatment of stellar evolution
  and chemical enrichment, as well as SN energy feedback in the form
  of galactic winds. As such, they can be used to extract the
  spectro--photometric properties of the simulated galaxies, which are
  identified as clumps in the distribution of star
  particles. Simulations have been carried out for a representative
  set of 19 cluster--sized halos, having mass $M_{200}$ in the range
  $5\times 10^{13}$--$1.8\times 10^{15}\msun$. All simulations have
  been performed for two choices of the stellar initial mass function
  (IMF), namely using a standard Salpeter IMF with power--law index
  $x=1.35$, and a top--heavy IMF with $x=0.95$. In general, we find
  that several of the observational properties of the galaxy
  population in nearby clusters are reproduced fairly well by
  simulations. A Salpeter IMF is successful in accounting for the
  slope and the normalization of the color--magnitude relation for the
  bulk of the galaxy population. In contrast, the top--heavy IMF
  produces too red galaxies, as a consequence of their exceedingly
  large metallicity. Simulated clusters have a relation between mass
  and optical luminosity which generally agrees with observations,
  both in normalization and slope. Also in keeping with
  observational results, galaxies are generally bluer, younger and
  more star forming in the cluster outskirts. However, we find that
  our simulated clusters have a total number of galaxies which is
  significantly smaller than the observed one, falling short by about
  a factor 2--3. We have verified that this problem does not have an
  obvious numerical origin, such as lack of mass and force
  resolution. Finally, the brightest cluster galaxies are always
  predicted to be too massive and too blue, when compared to
  observations. This is due to gas overcooling, which takes place in
  the core regions of simulated clusters, even in the presence of the
  rather efficient supernova feedback used in our simulations.
\end{abstract}


\section{Introduction} \label{sec:intro}

A key question in the study of the formation and evolution of galaxies
concerns the relationship between their observational properties and
the large--scale cosmological environment. In recent years, a
flourishing of observational campaign has provided a detailed
description of the evolution of the galaxy population in clusters. 
Indeed, galaxy clusters play a key role in the characterization of
the galaxy evolution. Each cluster provides a large sample of
galaxies, all placed at the same redshift. Furthermore, clusters offer
the possibility of sampling a variety of environments, from their
dense core regions, to the outskirts where the properties of the
cluster galaxy population tends to approach that of the field.

A diversity of the galaxy population in nearby clusters, with respect
to that in the field, was noticed already by
\cite{1974ApJ...194....1O} and by \cite{1980ApJ...236..351D}. Rich
clusters were shown to contain a higher fraction of bulge--dominated
(early type and S0) galaxies, and a correspondingly lower fraction of
star forming galaxies, than poor systems. \cite{1978ApJ...226..559B}
noticed that moderately distant clusters ($z\sim 0.3$) have a galaxy
population which is generally bluer than nearby clusters. This result
has been subsequently extended by a number of analyses and is
currently interpreted as high redshift clusters having larger
fractions of star--forming galaxies than local clusters \citep[see
][for a review]{2004bdmh.confE.104P}. \cite{1992MNRAS.254..601B} first
noticed that early--type members of the Coma cluster lie on a tight
relation in the color--magnitude diagram, a result that has been
subsequently confirmed by a number of analyses for extended samples of
nearby \citep[e.g., ][and references
therein]{1996A&A...309..749P,2003A&A...409...37A,2004ApJS..153..397B,2004ApJ...614..679L,2005ApJ...619..193M}
and distant clusters, out to $z\magcir 1$ \citep[e.g., ][and
references
therein]{2000ApJ...541...95V,2003ApJ...596L.143B,2004MNRAS.353..353A,2006astro.ph..1165S}.

As for the luminosity function (LF hereafter) of cluster galaxies,
much work has been done in recent years, with different groups
reaching different conclusions as for its universality, shape and
faint--end slope \citep[e.g., ][for a
review]{2004PASA...21..344D}. For instance, \cite{2006A&A...445...29P}
have recently analyzed the LF for sample of clusters identified in the
X--ray band within the ROSAT All Sky Survey (RASS, \citealt{VO92.1})
and covered by the Sloan Digital Sky Survey (SDSS
\citealt{2002AJ....123..485S}). They found no significant cluster to
cluster variations of the LF, once calculated within the same physical
radius ($r_{200}$ or $r_{500}$), with a shape fitted by a double
Schechter function to account for the upturn at faint
magnitudes. However, these results are at variance with respect to
those from other analyses \citep[e.g., ][]{2000A&A...353..930A}.

Within the widely accepted standard $\Lambda$CDM cosmological
scenario, galaxies arise from the hierarchical assembly of dark matter
(DM) halos. The gravitational dynamics of these halos is relatively
simple to describe to high precision with modern large supercomputer
simulations \citep[e.g.,][]{2005Natur.435..629S}. However, the
observational properties of galaxies are determined by the combined
action of the assembly of DM halos and by the physical processes which
define the evolution of the cosmic baryons. A complex interplay
between radiative gas cooling, star formation, chemical enrichment and
release of energy feedback from supernovae (SN) and active galactic
nuclei (AGN) is expected to determine the properties of the stellar
population in galaxies. At the same time, the cluster environment is
expected to play a significant role in altering the evolution of
galaxies. For instance, ram pressure exerted by the hot intra-cluster
medium (ICM) can lead to the removal of a substantial fraction of the
interstellar medium (ISM; \citealt{1972ApJ...176....1G}), thereby
affecting galaxy morphology, star formation and luminosity
\citep[e.g., ][, and references
therein]{1999MNRAS.308..947A,2004AJ....127.3361K}.

In this context, semi--analytical models of galaxy formation have been
used since several years as a flexible tool to study galaxy formation
within the cosmological hierarchical framework \citep[e.g., ][and
references
therein]{1993MNRAS.264..201K,1999MNRAS.310.1087S,2000MNRAS.319..168C,2002ApJ...575...18M}. A
powerful implementation of this method is that based on the so--called
hybrid approach, which combines N--body simulations, to accurately
trace the merging history of DM halos, and semi--analytic models to
describe the physics of the baryons \citep[e.g.,
][]{1999MNRAS.303..188K}.  \cite{SP01.2} applied this method to a DM
simulation of a cluster, with high enough resolution to allow them
resolving the population of dwarf galaxies. As a result, they found
that several observational properties (e.g., luminosity function,
mass-to-light ratio and morphological types) are rather well
reproduced. \cite{2001MNRAS.323..999D} applied a semi--analytical
model to a DM simulation of a large cosmological box, with the aim of
performing a combined study of kinematics, colors and morphologies for
both cluster and field galaxies. They concluded that a good agreement
with observations holds for cluster galaxies, while colors and star
formation rates of field galaxies were shown to evolve more rapidly
than observed. \cite{2006astro.ph..6591C} applied the same
semi--analytical model to a constrained simulation of the local
universe and concluded that significant differences exist between the
observed and the predicted properties of the large--scale distribution
of galaxy groups. \cite{2004MNRAS.349.1101D} incorporated in their
model also a scheme of metal production to follow the enrichment of
ICM and galaxies (see also \citealt{2006MNRAS.tmp..453C}). Among their
results, they found that the color--magnitude relation (CMR) is mainly
driven by metallicity effects, the redder galaxies on the sequence
being on average the more metal rich.  \cite{2005MNRAS.361..369L}
applied their semi--analytical model to a set of DM cluster
simulations. They also included a prescription to account for the
effect of ram--pressure stripping of the ISM as the galaxies move in
the hot cluster atmosphere, and found it to have only a very little
effect on the galaxy population.

A complementary approach to the semi--analytical models is represented
by using full hydrodynamical simulations, which include the processes
of gas cooling and star formation. The clear advantage of this
approach, with respect to semi--analytical models, is that galaxy
formation can be now described by following in detail the evolution of
the cosmic baryons while they follow the formation of the cosmic
web. However, the limitation of this approach lies in its high
computational cost, which prevents it to cover wide dynamic ranges and
to sample in detail the parameter space describing the processes of
star formation and feedback. For these reasons, describing the process
of galaxy formation with a self--consistent hydrodynamic approach
within the typical cosmological environment of $\sim 10\,$Mpc, relevant
for galaxy clusters, represents a challenging task for simulations of
the present generation.

In a pioneering paper, \cite{1994ApJ...437..564M} studied the effect
of including galaxies for the energy feedback and chemical enrichment
of the ICM. Since these simulations did not have enough resolution to
identify galaxies, they have been placed by hand, identifying them
with the peaks of the initial density
field. \cite{1996ApJ...472..460F} used for the first time a radiative
simulation of a cluster and identified galaxies as concentrations of
cooled gas. The aim of their study was to compare the dynamics of
member galaxies to that of DM particles. They concluded that galaxies
suffer for a substantial dynamical bias, a result which has not been
confirmed by more recent hydrodynamical simulations \citep[e.g.,
][]{2005MNRAS.358..139F,2006astro.ph..5151B}.  Thanks to the ever
improving supercomputing capabilities and efficiency of simulation
codes, a number of groups have recently completed hydrodynamical
simulations of galaxy clusters, which have good enough resolution to
trace the galaxy population with better
reliability. \cite{2005ApJ...618..557N} used simulations of eight
groups and clusters, performed with an adaptive mesh refinement
code, including star formation, feedback from supernovae and
chemical enrichment, to describe the spatial distribution of galaxies
inside clusters. They found that galaxies are more centrally
concentrated than DM sub--halos, with their number density profile
described by a NFW shape \citep{NA96.1}, although with a smaller
concentration parameter than for the DM
distribution. \cite{2005MNRAS.361..983R} used SPH simulations of two
clusters, including a similar physics, used spectrophotometric code to
derive galaxy luminosities in different bands. They analyzed the
resulting color--magnitude relations (CMR, hereafter) and luminosity
functions, claiming for an overall general agreement with
observations.

In this paper, we will present a detailed analysis of the galaxy
population for a set of 19 simulated clusters, which span the mass
range from $\simeq 5\times 10^{13}\msun$ to $\simeq 2\times
10^{15}\msun$. The simulations have been performed with the Tree-SPH
code {\small GADGET-2} \citep{2005MNRAS.364.1105S}. They include the
effect of radiative cooling, an effective model for star formation
from a multiphase ISM \citep{2003MNRAS.339..289S}, a phenomenological
recipe for galactic winds, a detailed stellar evolution model, thereby
accounting also for life--times and metal production from different
stellar populations (\citealt{2004MNRAS.349L..19T}; Tornatore et
al. in preparation). These simulations have been carried out also by
varying both the shape of the initial mass function (IMF, hereafter)
and the feedback strength. The inclusion of a detailed model of
chemical enrichment allows us to compute luminosities and colors for
galaxies of different metallicities, by using the GALAXEV
spectrophotometric code \citep{2003MNRAS.344.1000B}. In our analysis
we will concentrate on the properties of galaxy clusters at $z=0$. As
we shall discuss through the paper, several observational trends are
reproduced quite well by our simulations, although a number of
significant discrepancies are found. For this reason, the aim of our
analysis will be more that of understanding the directions to improve
simulations, rather than seeking for a best fitting between model
predictions and observations.

The plan of the paper is as follows. In Section 2 we provide the
general characteristics of the simulated clusters and describe the
relevant features of the {\small GADGET-2} version used for this
analysis. In Section 3 we will describe the method of galaxy
identification and how luminosities in different bands are
computed. Section 4 contains the description of the properties of the
simulated galaxy population and their comparison with observational
data. In particular, we will discuss the radial distribution of
galaxies, the CMR, the mass--luminosity ratio, the luminosity
function, the star formation rate and the color and age gradients. Our
main results will be summarized and discussed in Section 5. We will
discuss in an Appendix the effects of numerical resolution on the
stability of the results of our analysis.

\section{The simulations} 
\label{sec:sims}
\subsection{The simulated clusters}
\label{s:simul}
Our set of clusters are identified within nine Lagrangian regions,
centered around as many main clusters. They were extracted from a
DM--only simulation with a box size of $479\,h^{-1}$Mpc of a flat
$\Lambda$CDM model with $\Omega_m=0.3$ for the matter density
parameter, $h=0.7$ for the Hubble constant in units of 100 km
s$^{-1}$Mpc$^{-1}$, $\sigma_8=0.9$ for the r.m.s. fluctuation within a
top--hat sphere of $8\hm$ radius and $\Omega_{\rm b}=0.04$ for the
baryon density parameter \citep{2001MNRAS.328..669Y}. 

Thanks to the fairly large size chosen for these Lagrangian regions,
several of them contain other interesting clusters, besides the main
one. In this way, we end up with 19 clusters with mass $M_{200}$ in
the range\footnote{We define $M_{\Delta}$ as the mass contained within
  a radius encompassing a mean density equal to $\Delta \rho_c$, with
  $\rho_c$ the critical cosmic density.}  $5\times 10^{13}-1.8\times
10^{15}\msun$, out of which 4 clusters have $M_{200}>10^{15}\msun$
(see Table \ref{t:clus}). Mass resolution is increased inside the
interesting regions by using the Zoomed Initial Condition (ZIC)
technique by \cite{1997MNRAS.286..865T}. Unperturbed particles
positions were placed on a `glass' \citep{WH96.1b}, and initial
displacements were then assigned according to the Zeldovich
approximation \citep[e.g.][]{1989RvMP...61..185S}.  Besides the
low--frequency modes, which were taken from the initial conditions of
the parent simulation, the contribution of the newly sampled
high--frequency modes was also added.  The mass resolution was
progressively degraded in more distant regions, so as to save
computational resources while still correctly describing the
large--scale tidal field of the cosmological environment.

\begin{table}
\centering
\caption{Characteristics of the clusters identified within the
  simulated regions at $z=0$. Col. 1: name of the simulated region;
  Col. 2: name of the clusters within each region; 
  Col. 3: value of the total
  mass, $M_{200}$, contained within the radius $r_{200}$ encompassing an average
  density 200 times larger than the critical cosmic density $\rho_c$
  (units of 
  $10^{14}\msun$); Col. 4: total number of galaxies, $N_{200}$, within
  $r_{200}$, having a minimum number of 32 star particles.}
\begin{tabular}{llcr}
Region name & Cluster name & $M_{200}$ & $N_{200}$ \\ 
\hline 
g1          & g1.a         & 12.9     &  418 \\ 
            & g1.b         & 3.55     &  149 \\
            & g1.c         & 1.39     &   51 \\
            & g1.d         & 0.96     &   33 \\
            & g1.e         & 0.64     &   35 \\
g8          & g8.a         & 18.4     &  589 \\
            & g8.b         & 1.02     &   42 \\
            & g8.c         & 0.67     &   18 \\
            & g8.d         & 0.59     &   26 \\
            & g8.e         & 0.54     &   21 \\
g51         & g51.a        & 10.9     &  371 \\
g72         & g72.a        & 10.7     &  440 \\
            & g72.b        & 1.55     &   60 \\
g676 & g676.a & 0.89 & 23 \\
g914 & g914 a & 0.86 & 16 \\
g1542& g1542.a& 0.89 & 34 \\
g3344& g3344.a& 0.97 & 29 \\
g6212& g6212.a& 0.92 & 22 \\
\end{tabular}
\label{t:clus}
\end{table}

Once initial conditions are created, we split particles in the
high--resolution region into a DM and a gas component, whose mass
ratio is set to reproduce the assumed cosmic baryon fraction. Instead
of placing them on top of each other, in order to avoid spurious
numerical effects, we displace gas and DM particles such that the
centre of mass of each parent particle is preserved and the final gas
and dark matter particle distributions are interleaved by one mean
particle spacing.  In the high--resolution region, the masses of the
DM and gas particles are set to $m_{\rm DM}=1.13\times
10^9\,h^{-1}{\rm M}_\odot$ and $m_{\rm gas}=1.7\times 10^8\,h^{-1}{\rm
M}_\odot$, respectively. The Plummer--equivalent softening length for
the gravitational force is set to $\epsilon_{\rm Pl}=5.0\, h^{-1}$kpc,
kept fixed in physical units from $z=5$ to $z=0$, while being
$\epsilon_{\rm Pl}=30.0\, h^{-1}$kpc in comoving units at higher
redshift.

\subsection{The code}
\label{s:code}
Our simulations are based on an evolution of {\small
  GADGET-2}\footnote{http://www.MPA-Garching.MPG.DE/gadget/}
\citep{SP01.1,2005MNRAS.364.1105S}, which includes a detailed
treatment of chemical enrichment from stellar evolution
(\citealt{2004MNRAS.349L..19T}; Tornatore et al., in
preparation). {\small GADGET-2} is a parallel Tree+SPH code with fully
adaptive time--stepping, which includes an integration scheme which
explicitly conserves energy and entropy \citep{2002MNRAS.333..649S},
radiative cooling, the effect of a uniform and evolving UV background
\citep{1996ApJ...461...20H}, star formation from a multiphase
interstellar medium and a prescription for galactic winds triggered by
SN explosions (see \citealt{2003MNRAS.339..289S} for a detailed
description, SH03 hereafter), and a numerical scheme to suppress
  artificial viscosity far from the shock regions (see
  \cite{2005MNRAS.364..753D}). In the original version of the code,
energy feedback and global metallicity were produced only by SNII
under the instantaneous--recycling approximation (IRA).

We have suitably modified the simulation code, so as to correctly
account for the life--times of different stellar populations, to
follow metal production from both SNIa and II, while
self--consistently introducing the dependence of the cooling function
on metallicity by using the tables by \cite{1993ApJS...88..253S}. A
detailed description of the implementation of these algorithms will be
presented in a forthcoming paper (Tornatore et al., in preparation),
while we provide here a short descriptions of the most relevant
features of the code.

In order to maintain the general approach of the multiphase model by
SH03, we assume that stars with masses $>40\,M_\odot$ explode into
SNII soon after their formation, thereby promptly releasing energy and
metals. In contrast, we correctly account for the lifetime of stars
having masses smaller than $40\,M_\odot$. The simulations that we will
discuss here use the lifetimes provided by \cite{1989A&A...210..155M},
which have been shown to reproduce the abundance pattern in the Milky
Way \citep{1997ApJ...477..765C}. Within the stochastic approach to
star formation (SH03), each star particle is generated with a mass
equal to one third of the mass of its parent gas particle.

Therefore, each star particle is considered as a single stellar
population (SSP, hereafter), with its own mass, metallicity and
redshift of formation. For each SSP we compute both the number of
stars turning into SNII and Ia at each time-step and the number of
stars ending their AGB phase. Then we calculate the amount of energy
and metals produced by each star particle in a given time interval,
decreasing accordingly the mass of the particle.  In this way, each
star particle is characterized by both its initial mass, assigned at
the time of its formation, and its final mass, which is updated during
the evolution.  Both SNII and SNIa are assumed to release $10^{51}$
ergs each, while no energy output is associated to the mass loss from
AGB stars. The relative number of SNII and SNIa depends on the choice
of the stellar initial mass function (IMF). In the following, we will
assume for the IMF the power--law shape $dN/d\log{m}\propto
m^{-x}$. Simulations will be run by assuming the Salpeter IMF with
$x=1.35$ \citep[][Sa-IMF hereafter]{1955ApJ...121..161S} and a
top--heavy IMF with $x=0.95$ (\citealt{1987A&A...173...23A}, TH-IMF
hereafter).

The SNIa are associated to binary systems whose components are in the
0.8--$8\,M_\odot$ mass range \citep{1983A&A...118..217G}, while SNII
arise from stars with mass $>8\,M_\odot$. In the following, we will
assume that 10 per cent of stars in the 0.8--$8\,M_\odot$ mass range
belongs to binary systems, which then produces SNIa. We use the
analytical fitting formulas for stellar yields of SNIa, SNII and PNe
provided by \cite{2001MNRAS.322..800R}, and based on the original
nucleosynthesis computations of \cite{1997NuPhA.621..467N}, using
their W7 model, \cite{1995ApJS..101..181W} and
\cite{1981A&A....94..175R}. The formulation for the SNIa rate has been
calculated as in \cite{2001ApJ...558..351M}. In the simulations that
we present, besides H and He, we have followed Fe, O, C, Si, Mg,
S. Once produced by a star particle, metals are spread over the same
number of neighbours, 64, used for the SPH computations, also using
the same kernel. We normalize the IMFs in the mass range 0.1--100
$M_\odot$. Owing to the uncertainty in modelling yields for very
massive stars, we take yields to be independent of mass above
$40\,M_\odot$. While any uncertainty in the yields of such massive
stars has a negligible effect for the Salpeter IMF, their accurate
description \citep[e.g.,][]{1996ApJ...460..408T,2002ApJ...567..532H}
is required when using a top--heavier IMF.

Our prescription to account for stellar evolution in the simulations
implies a substantial change of the multiphase ``effective model'' by
SH03, which we have suitably modified to account for {\em (a)} the
contribution of the energy reservoir provided by the SN which are
treated outside the IRA, and {\em (b)} the metal--dependence of the
cooling function, that we introduce using the tables from
\cite{1993ApJS...88..253S}. The resulting density threshold for a gas
particle to become multiphase, thereby being eligible to undergo star
formation, is fixed to $n_H=0.1$ cm$^{-3}$ at zero
metallicity. According to eq.(23) of SH03, this threshold is inversely
proportional to the cooling function. Since the latter depends on
metallicity, we take self--consistently a metallicity--dependent
star--formation threshold for gas having a non-zero metallicity.

SH03 also provided a phenomenological description for galactic winds,
which are triggered by SN energy release and whose strength is
regulated by two parameters. The first one gives the wind mass loading
according to the relation, $\dot M_W=\eta \dot
M_*$, where $\dot M_*$ is the star formation rate. Following SH03, we 
assume $\eta=3$. The second parameter is the wind velocity, $v_w$.
For the runs based on the Salpeter IMF, we always use
$v_w=500\vel$. For the above values of $\eta$ and $v_w$, all the
energy from SNII is converted in kinetic energy, as in the original
SH03 paper. \cite{2003MNRAS.339..312S} made a study of the star
formation history predicted by hydrodinamical simulations which
include galactic winds with a similar velocity. They concluded that
the resulting star fraction at $z=0$ and high--$z$
star formation history are comparable to the observed ones. In order
to verify the effect of galactic ejecta, the g676 and g51 regions are also
simulated with the Salpeter IMF, but setting to zero the wind velocity
(Sa-NW runs). As for
the runs based on the top--heavy IMF, we will also use $v_W=500\vel$
for all clusters, with the exception of g676 and g51 regions, for
which we also use $v_w=1000\vel$ (TH-SW runs). In this case, the two
wind speeds correspond to an energy budget of about 0.4 and 1.4 times
the energy provided by SNII.
An efficiency larger than unity can be justified on the ground of the
large uncertainties on the actual energy released by SNII
explosions. In this perspective, we also take the value of the wind
velocity as a confidence value. A wind velocity $v_W=1000\vel$ is
intended to represent an extreme feedback case, so that comparing the
results with the two values of wind velocity allows us to check the
effect of a stronger feedback on the final properties of the galaxy
population. We summarize in Table \ref{t:sim} the IMFs and feedback
used in our simulations.

\begin{table}
\centering
\caption{Characteristics of the simulations. Col. 1: simulation name;
  Col. 2: IMF slope; Col 3: wind speed, $v_{w}$ (units of $\vel$).} 
\begin{tabular}{ccc}
Name & IMF slope & $v_{w}$ \\ 
\hline 
Sa     & 1.35    & 500 \\
Sa-NW  & 1.35    & 0 \\
TH     & 0.95    & 500 \\
TH-SW  & 0.95    &1000 \\ 
\end{tabular}
\label{t:sim}
\end{table}

\begin{figure*}
\centerline{
\hbox{
\psfig{file=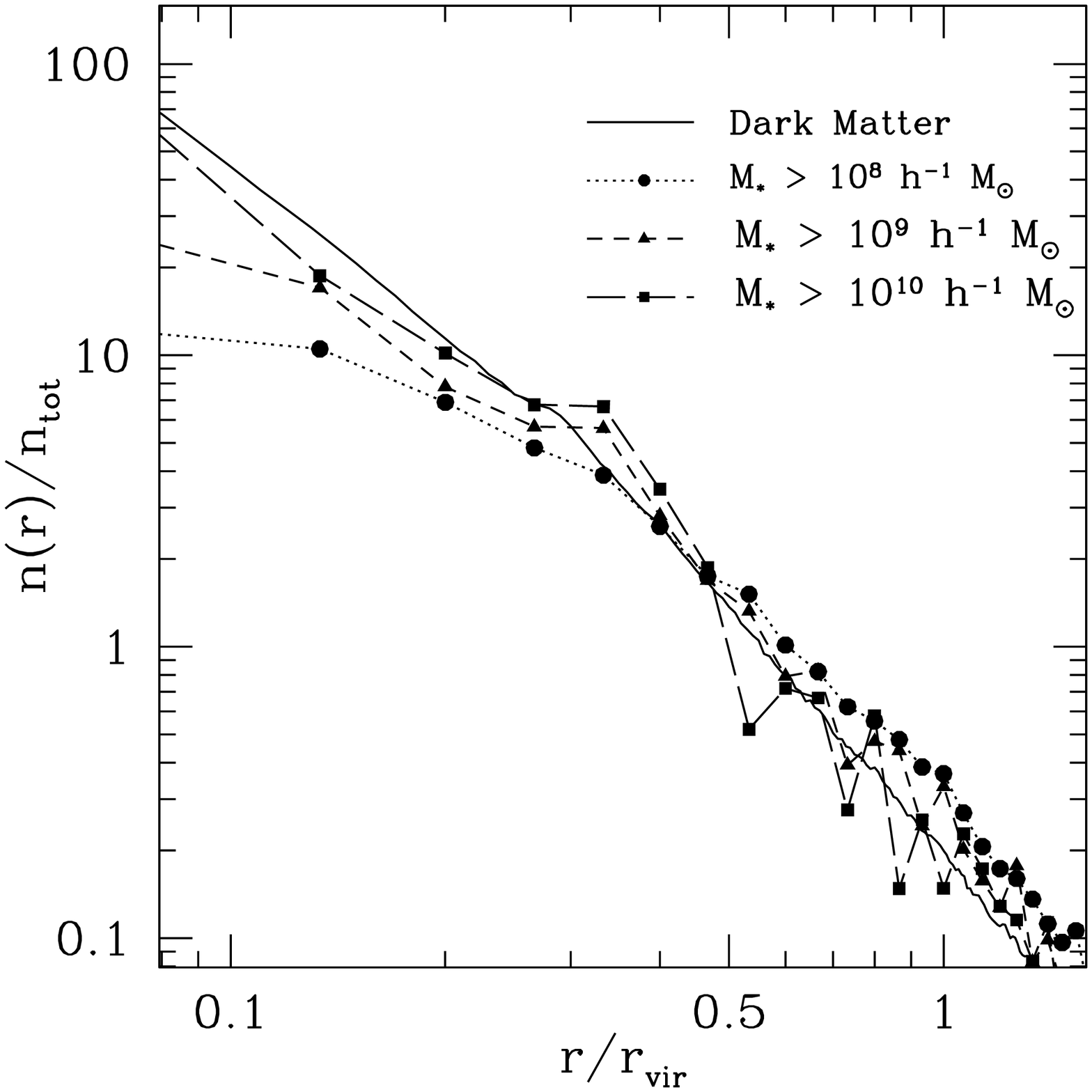,width=7.5cm} 
\psfig{file=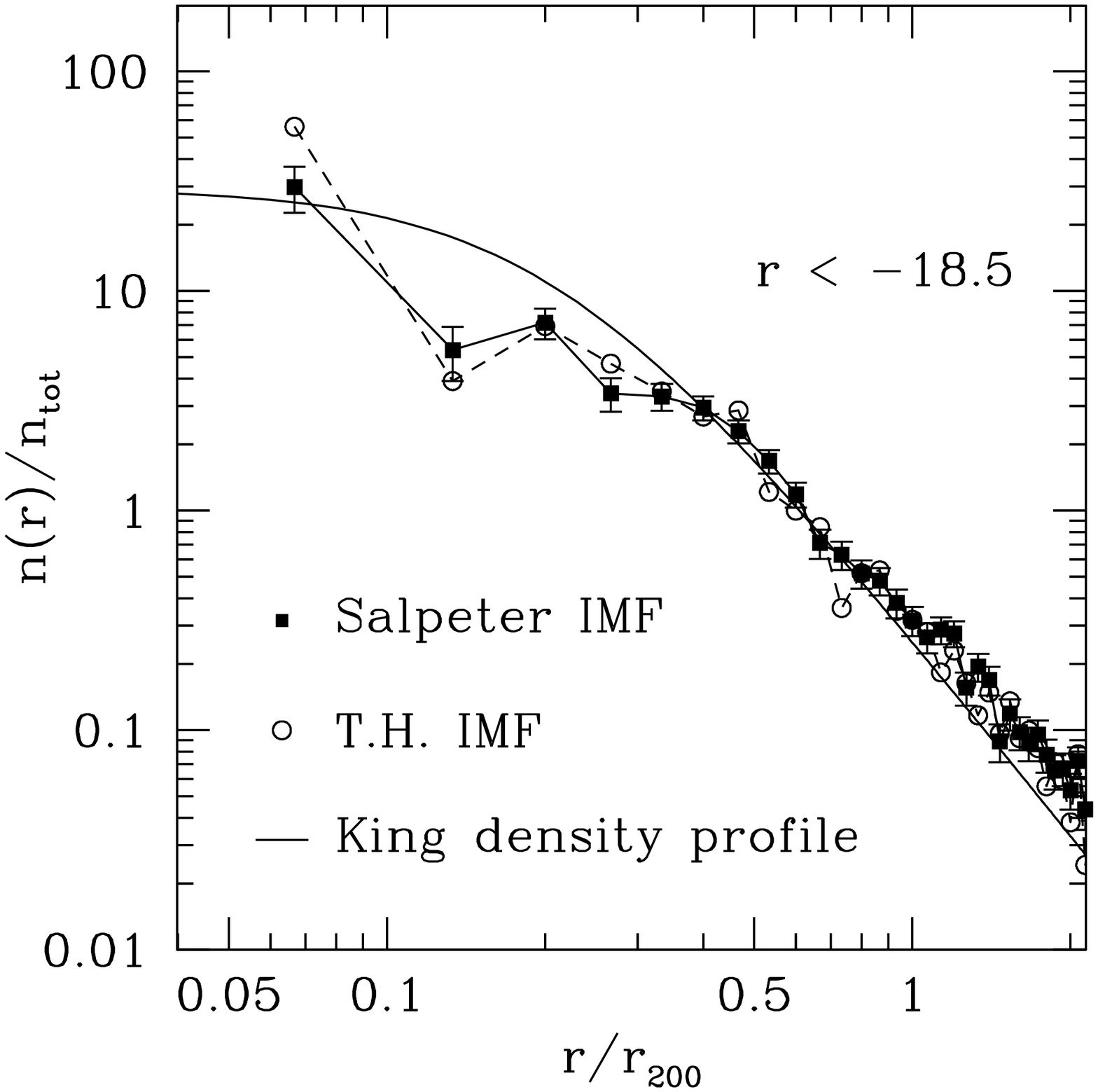,width=7.5cm} }}
\caption{The number density profile of cluster galaxies. Left panel:
  the profiles of galaxies of different stellar mass, averaged over
  all clusters, for the runs with Salpeter IMF (filled symbols). Shown
  with the solid curve is the average DM profile. All profiles are
  normalized to the total number density within the virial radius.
  Right panel: The number density of galaxies, brighter than
  $r=-18.5$, contained within a given radius, normalized to the total
  number density found within the $r_{200}$. Filled squares and open
  circles are by combining all the simulated clusters, for the
  Salpeter and for the top-heavy IMF, respectively. The solid curve is
  the best--fit King model to the number density profiles of cluster
  galaxies from the analysis of RASS--SDSS data by
  \protect\cite{2006astro.ph..6260P}, here plotted with arbitrary
  normalization. Errorbars in the simulation profile correspond
  to Poissonian uncertainties. For reasons of clarity they have been
  plotted only for the Salpeter runs.}
\label{fi:densprof} 
\end{figure*}

\section{Assigning luminosities to galaxies}
\label{s:lumi}
As a first step, we identify galaxies from the distribution of star
particles by applying the SKID algorithm\footnote{See {\tt
http://www-hpcc.astro.washington.edu/tools/ skid.html} }
\citep{2001PhDT........21S}. We provide a short description of how we
applied this algorithm, while a more detailed discussion and
presentation of tests is provided elsewhere (Murante et al. 2006, in
preparation; see also \citealt{2006MNRAS.367.1641B}).  An overall
density field is computed by using the distribution of all the
particle species, by using a SPH spline--kernel.  The star particles
are then moved along the gradient of the density field in steps of
$\tau/2$, where we assume $\tau\simeq 3 \epsilon_{\rm Pl}$. When a
particle begins to oscillate inside a sphere of radius $\tau/2$, it is
stopped.  Once all particles have been moved, they are grouped using a
friends-of-friends (FOF) algorithm, with linking length $\tau/2$,
applied to the new particle positions.  The binding energy of each
group identified in this way is then used to remove from the group all
star particles which are recognized as unbound. All particles in a
sphere of radius $\tau$, centered on the center of mass of the group,
are used to compute such a gravitational binding energy. Finally,
we identify as ``bona fide'' galaxies only those SKID--groups
containing at least 32 star particles after the removal of unbound
stars.

Since each star particle is treated as a SSP, with formation redshift
$z_f$ and metallicity $Z$,we can assign to it luminosities in
different bands by resorting to a spectrophotometric code, for the
appropriate IMF used in the corresponding simulation.

To this purpose, we have used the outputs of the GALAXEV code
\cite{2003MNRAS.344.1000B} to create a grid of metallicity and age
values for a SSP of $1M_\odot$. Luminosities in different bands are
then assigned to each SSP of this grid. Since two different IMFs are
used for our simulations, this grid is also computed for both the
\cite{1955ApJ...121..161S} and the top--heavy IMF. Note that GALAXEV
assumes that contributions of different metal species to the total
metallicity are in solar proportions, while this is not necessarily
true for the star particles in our simulations. For this reason, we
use the total metallicity of each star particle (i.e., the sum of the
contributions from the different elements) as input for GALAXEV. We
have verified that, using instead Iron or Oxygen as proxy for the
global metallicity, our final results are left essentially
unchanged. Consistent with the stellar evolution implemented in the
simulation code, GALAXEV accounts for stellar mass loss. Therefore, we
use the initial mass of each star particle in the simulations as input
to GALAXEV to compute the corresponding luminosities. For each star
particle we interpolate its age and metallicity with the appropriate
entries of the grid. Finally, we evaluate the luminosity $L_{\star
,\nu}$ of each star particle, which is treated as a SSP of mass
$M_\star$ and age $t$, in the $\nu$ band by:
\begin{equation}
L_{\star,\nu}(t)\,=\,{M_{\star}(t)\over M\odot}\,L_\nu(1M\odot)
\end{equation}
In this way, the luminosity in the $\nu$ band of each galaxy is given
by the sum of the luminosities contributed by each member star
particle. As a final result, for each galaxy our analysis provides
stellar mass, mean stellar age, metallicity, star-formation rate
(SFR), absolute magnitudes in the $U,B,V,R,I,J,K$, bolometric standard
Johnson bands and in the $g,u,r,i,z$ SLOAN bands.

We note that GALAXEV accounts for metallicity values in the range
0.005--2.6 $Z_\odot$. While only a negligible number of stars have a
metallicity below the lower limit of this interval, a sizeable number
of particles, especially for the top--heavy IMF, are found with metallicities
exceeding the upper limit. Whenever the particle metallicity lies
outside the above range, they are set to the value of the nearest
boundary.

\section{Results}

\subsection{The number-density profile of cluster galaxies}
\label{s:numb}
A well established result from collisionless simulations of galaxy
clusters is that the radial distribution of galaxy--sized subhalos is
less concentrated than that of DM
\citep[e.g.][]{2000ApJ...544..616G,SP01.2,2004MNRAS.348..333D}, and
also less concentrated and more extended than the observed radial
distribution of cluster galaxies
\citep[e.g.][]{2004MNRAS.352..535D,2004MNRAS.352L...1G}. While some
residual numerical overmerging can still be present at the high
resolution achieved in DM--only simulations, it has been suggested
\citep{2004MNRAS.352..535D} that overmerging may be physical in origin
and related to the dissipationless dynamics. The possibility to
include radiative cooling and star formation in hydrodynamical
simulations allows one to verify whether the same result holds also
for the galaxies identified from the star distribution. The general
conclusion from these analyses is that the radial distribution of
simulated galaxies is indeed more concentrated than that of DM
sub-halos \citep[e.g.,][]{2005ApJ...618..557N}. An intermediate
approach, based on coupling semi--analytical models of galaxy
formation with high resolution collisionless simulations
\citep[e.g.,][]{SP01.2,2004ApJ...609...35K,2004MNRAS.352L...1G,2004MNRAS.349.1101D,2005MNRAS.361..369L},
confirms that the radial distribution of galaxies is more extended
compared to DM. Clearly, the dynamics of halo formation in these
studies is driven by the collisionless component. Therefore, suitable
effective recipes should be included to prevent physical overmerging
of galaxies within DM halos, so as to achieve agreement with the
observed radial galaxy distribution
\citep[e.g.][]{SP01.2,2004MNRAS.349.1101D}.

We show in the left panel of Figure \ref{fi:densprof} the number
density profiles of cluster galaxies, after averaging over all the
simulated clusters, compared to the corresponding average DM density
profile. As discussed by \cite{2005ApJ...618..557N}, selecting
galaxies in hydrodynamical simulations of clusters, which include star
formation, produces profiles which are generally steeper than those of
DM halos. We confirm here that galaxy profiles become closer to the DM
profile as more massive galaxies are selected, with objects more
massive than $10^{10}\msun$ tracing a distribution quite close to the
DM one. This is just the consequence of the improved capability of
more massive galaxies to preserve their identity within merging
halos. Indeed, since galaxies are more concentrated than their hosting
DM halos, they are able to better survive to disruption and merging,
thereby providing a better sampling of the underlying DM
distribution. While this trend is generally consistent with
observations, a close comparison with data requires assigning
luminosities to simulated galaxies. For this reason, we also show in
the right panel of Fig.\ref{fi:densprof} the number density profiles
of galaxies brighter that $r=-18.5$ and compare it to the best fit to
the observed profiles by \cite{2006astro.ph..6260P}, who trace these
profiles out to $\simeq 2r_{200}$. Galaxies of this luminosity have a
typical stellar mass of the order of $5\times 10^9 M_\odot$ in the
simulations with Salpeter IMF. In general, the observed and the
simulated profiles are quite similar down to $\simeq 0.4 r_{200}$. At
smaller radii there is a trend for simulated galaxies to have a lower
number density than in real clusters. Therefore, although tracing
galaxies instead of DM halos helps in the comparison with the observed
galaxy profiles, still overmerging is also the likely reason for the
shallower profile as traced by simulated galaxies. 

\begin{figure}
\centerline{
\psfig{file=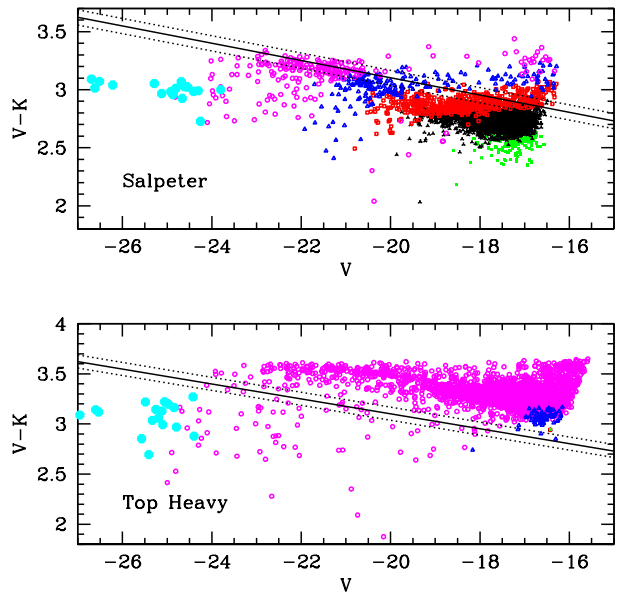,width=8.5cm} 
}
\caption{The $V$--$K$ vs. $V$ color--magnitude relation by combining
  all the galaxies within the virial radii of the simulated clusters,
  for the Salpeter IMF (top panel) and for the top--heavy IMF with
  normal feedback (bottom panel). Straight lines in each panel show
  the observed CMR relations by \protect\cite{1992MNRAS.254..601B}, with the
  corresponding intrinsic standard deviations. Big filled dots mark
  the BCG of each cluster. Different symbols and colors are used for
  galaxies having different metallicities. Magenta open circles:
  $Z>1.5Z_\odot$; blue filled triangles: $1.5<Z/Z_\odot<1$; red open
  squares: $1<Z/Z_\odot<0.7$; black open triangles:
  $0.7<Z/Z_\odot<0.4$; green filled squares: $Z<0.4Z_\odot$.}
\label{fi:cmr_z0}
\end{figure}

\begin{figure}
\centerline{
\psfig{file=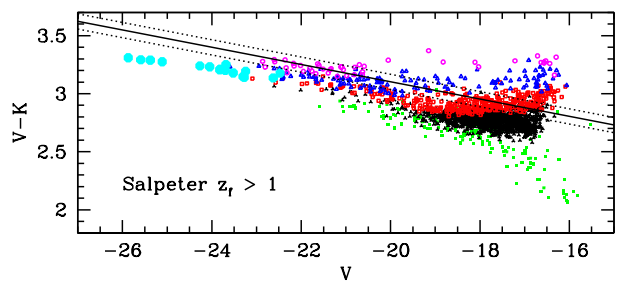,width=8.5cm} 
}
\vspace{-4.5truecm}
\caption{The same as the top panel of Figure \ref{fi:cmr_z0}, but
  only including in the computation of the luminosities the star
  particles having redshift of formation $z_f>1$.}
\label{fi:cmr_z1}
\end{figure}

\subsection{The color--magnitude relation}
\label{s:cmr}
Bright massive ellipticals, which dominate the population of cluster
galaxies, are observed to form a tight correlation between galaxy
colors and magnitudes, the so--called red sequence or color--magnitude
relation (CMR)
\citep[e.g.,][]{1992MNRAS.254..601B,1996A&A...309..749P,2004MNRAS.353..353A,2004ApJ...614..679L,2005ApJS..157....1G,2005ApJ...619..193M}.
Attempts to compare the observed CMR to that predicted by cosmological
models of galaxy formation have been attempted both using
semi--analytical approaches
\citep[e.g.,][]{2004MNRAS.349.1101D,2005MNRAS.361..369L} and full
hydrodynamical simulations \citep{2005MNRAS.361..983R}. As a general
results, model predictions reproduce the slope of the CMR reasonably
well, but with a scatter which is generally larger than
observed. \cite{2005MNRAS.361..983R} simulated one Virgo--sized and
one Coma--sized cluster. They found that a top--heavy IMF reproduces
the normalization of the CMR better than a Salpeter IMF, which gives
too blue colors as a consequence of the too low metallicity. This
conclusion is at variance with respect to that reached by
\cite{2004MNRAS.348..333D} who, instead, reproduce the correct CMR
normalization with a Salpeter IMF.

\begin{figure*}
\centerline{
\hbox{
\psfig{file=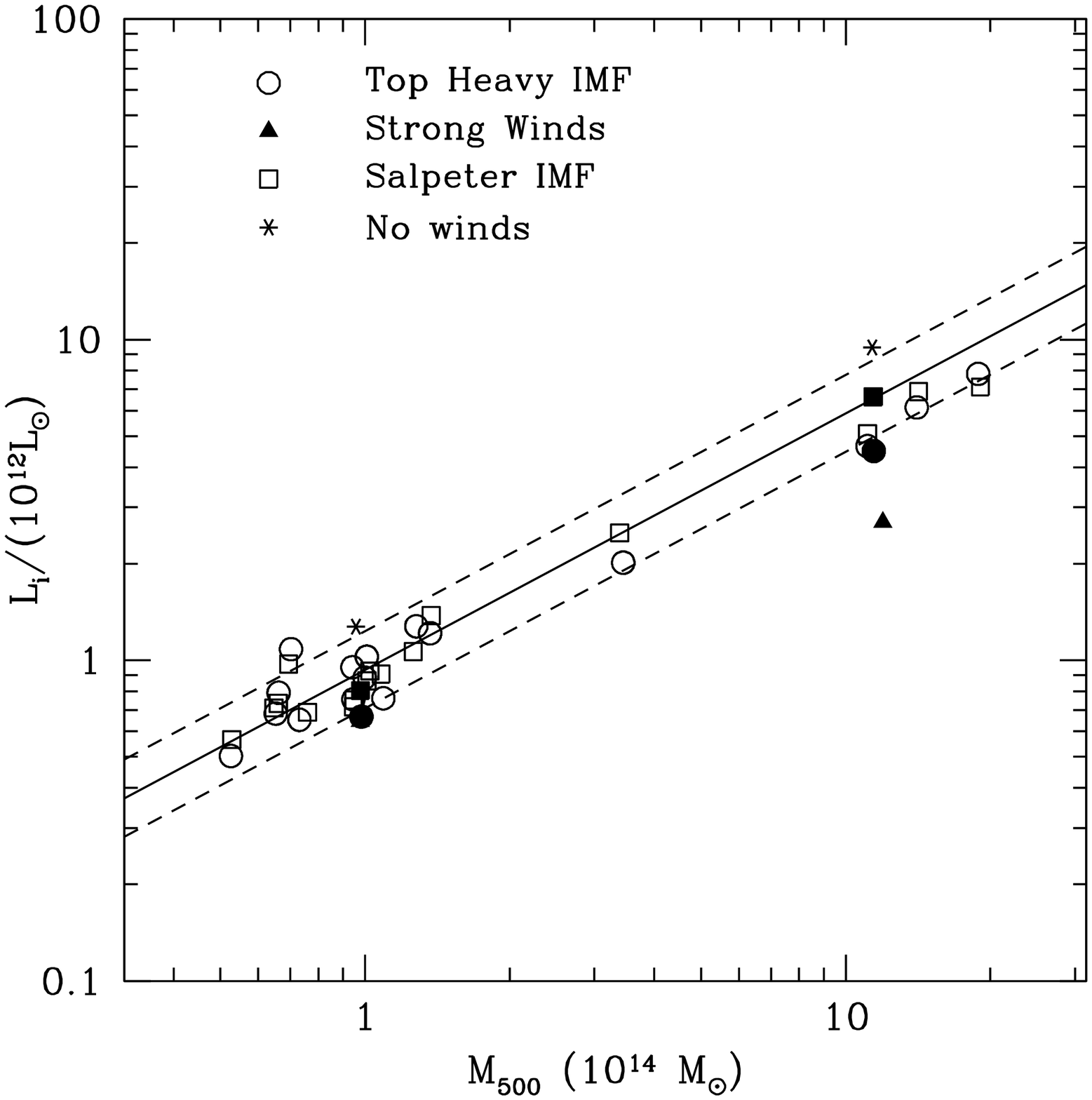,width=7.5cm} 
\psfig{file=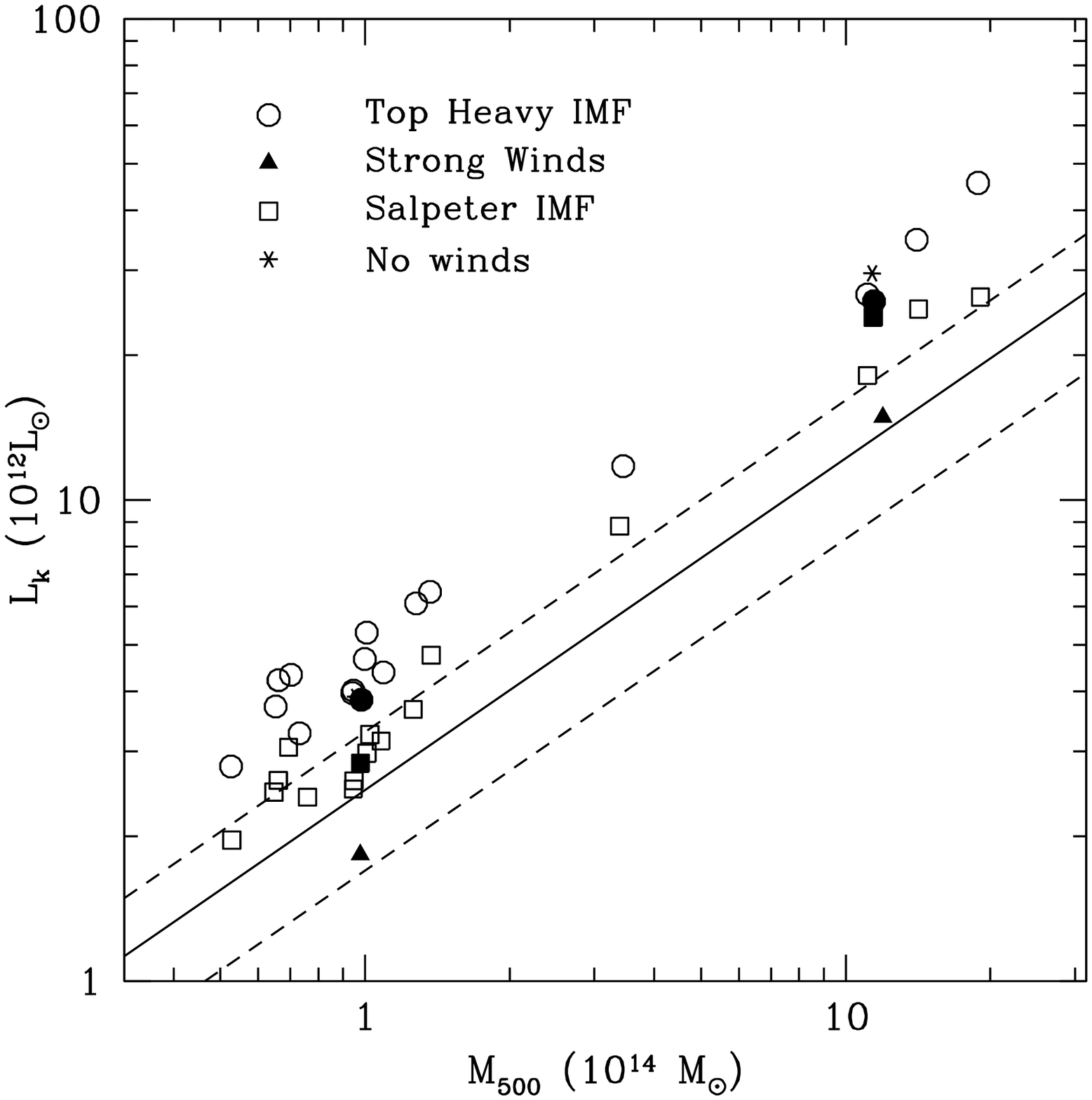,width=7.5cm} 
}}
\caption{The comparison between simulated and observed relation
  between mass and luminosity in the $i$ band (left panel) and in the
  K band (right panel). In each panel, 
  squares are for the Salpeter IMF, circles for the top--heavy IMF
  with normal feedback, triangles for the top--heavy IMF with strong
  feedback and asterisks for the Salpeter IMF with no feedback. Filled
  squares and circles are for the g676 and g51 runs, so as to
  make clear the effect of changing the feedback strength. 
  \protect\cite{2005A&A...433..431P} for the $i$ band and from
  \protect\cite{2004ApJ...610..745L} for the $K$ band, with the dashed lines
  marking the corresponding observational scatter.}
\label{fi:ml} 
\end{figure*}

Thanks to the larger number of simulated clusters, we can perform the
comparison between simulated and observed CMR with much improved
statistics. The results of this comparison are shown in Figure
\ref{fi:cmr_z0}. As a term of comparison, we use the observational
determinations by \cite{1992MNRAS.254..601B} (see also
\citealt{2001MNRAS.326.1547T}) for the $V-K$ vs. $V$ CMR, which has
been determined in the magnitude range $[-18,-23]$. Quite apparently,
a Salpeter IMF is successful in reproducing the correct amplitude of
the CMR at the bright end, $V\mincir -20$, while it tends to produce
too blue faint galaxies. The slope of the CMR appears to be driven by
metallicity, the brighter galaxies being redder mainly as a
consequence of their higher metal content, thus in line with the
interpretation by \cite{1997A&A...320...41K}. 
At the same time, we note that a top--heavy IMF produces too
metal--rich galaxies, thereby inducing too high a normalization of the
CMR.

The metal content of galaxies is clearly determined by the combined
action of stellar nucleosynthesis and other processes which bring
enriched gas far from star forming regions, thus preventing all metals
from being locked back in newly forming stars. Processes, such as ram
pressure stripping \citep[e.g., ][]{2005astro.ph..7605D} and galactic
winds \citep[e.g., ][]{2001ApJ...560..599A} have been suggested as the
possible mechanisms to enrich the diffuse intergalactic
medium. Clearly, the more efficient these mechanisms, the lower the
expected metallicity of stars and, therefore, the bluer their colors.
In order to verify whether more efficient galactic winds may decrease
the metallicity of galaxies in the runs with top--heavy IMF, we have
re--simulated the g676 and g51 clusters using $v_w=1000 \vel$ for the
galactic winds (TH-SW runs). However, while the effect of the stronger
feedback is that of decreasing the number of galaxies, (see also
Sect. \ref{s:lf} below), it leaves their metal content, and,
therefore, the high CMR normalization, almost unchanged.

Although a Salpeter IMF fares rather well as for the CMR, we
note that all the BCGs (big filled circles in Fig.\ref{fi:cmr_z0}) are
much bluer, by about 0.5 magnitude, than expected from the red
sequence. Such a blue excess of the colors of the BGCs, which takes
place despite their high metallicity, finds its origin in the large
star formation rate, associated to overcooling, which takes place in
the central cluster regions. Typical values for the star formation
rate of the BCG in our simulations are in range 600--1000 $M_\odot$/yr
for the most massive clusters ($M_{200}\simeq 10^{15}\msun$) and $\sim
100 M_\odot$/yr for the least massive ones ($M_{200}\simeq
10^{14}\msun$). Although observations indicate the presence of some
ongoing star formation in some BCGs located at the center of cool core
clusters, they are always at a much lower level and consistent with a
star formation rate of $\sim 10$--$100 M_\odot$/yr for clusters of
comparable richness \citep[e.g., ][and references
therein]{1987MNRAS.224...75J,2006astro.ph..2323B,2006astro.ph..4044M}.

  The effect of recent star formation on the CMR is explicitely
  shown in Figure \ref{fi:cmr_z1}. We show here the case in which all
  star particles, formed at redshift $z<1$ are excluded from the
  computation of the galaxy luminosities. This is equivalent to assume that
  we completely quench star formation since $z=1$. Neglecting recent
  star formation has the twofold effect of reducing the scatter in the
  CMR and of making BCG colors significantly redder, although they
  still fall slightly below the observed relation.

\subsection{The mass--luminosity ratio}
\label{s:ml}
A number of observational analyses have established that the
mass--to--light ratio in clusters generally increases with the cluster
mass, $M/L\propto M^\gamma$ with $\gamma \simeq 0.2$--0.4, over a
fairly large dynamic range, from poor groups to rich clusters
\citep[e.g.,
][]{1998A&A...331..493A,2000ApJ...530...62G,2002ApJ...569..720G,2002ApJ...565L...5B,2003ApJ...591..749L,2004ApJ...610..745L,2004AJ....128.1078R,2004AJ....128.2022R,2005A&A...433..431P}.
A likely explanation for this trend is the reduced cooling efficiency
within more massive, hotter halos \citep[e.g.,
][]{2003MNRAS.339..312S}, which reduces star formation within richer
clusters. In fact, an increasing trend of $M/L$ with cluster mass is
naturally predicted by semi--analytical models of galaxy formation
\citep[e.g., ][]{1999MNRAS.303..188K}.

In Figure \ref{fi:ml} we compare the relation between mass and
luminosity within $r_{500}$ for our simulated clusters, and compare it
to the $i$--band results by \cite{2005A&A...433..431P} and to the
$K$--band results by \cite{2004ApJ...610..745L}. In general, we find
that the $M/L$ from simulations is rather close to the observed one in
the $i$ band, also with a comparably small scatter. In the $K$ band, a
Salpeter IMF still agrees with observations within the statistical
uncertainties, while the top--heavy IMF produces too red galaxies,
thus consistent with the results of the CMR, as shown in
Fig.\ref{fi:cmr_z0}. We fit our mass--luminosity relation with a
power--law

\be 
{L\over 10^{12}L_\odot}\,=\,\beta\left({M_{500}\over
  10^{14}M_\odot}\right)^\alpha \,,
\ee

we find $(\alpha,\beta)_i=(0.74,0.92)$ and $(\alpha,\beta)_K=(0.76,3.2)$
in the $i$ and $K$ band, respectively, for the runs with Salpeter IMF,
while $(\alpha,\beta)_i=(0.70,0.91)$ and $(\alpha,\beta)_K=(0.74,4.7)$
for the top--heavy IMF. Therefore, our simulations agree with the
observational trend for an increasing mass-to-light ratio with cluster
mass, independent of the IMF and luminosity band.
Using the stronger feedback for the top--heavy IMF turns into
a sizeable suppression of the luminosity, especially for g51.

\begin{figure*}
\centerline{
\hbox{
\psfig{file=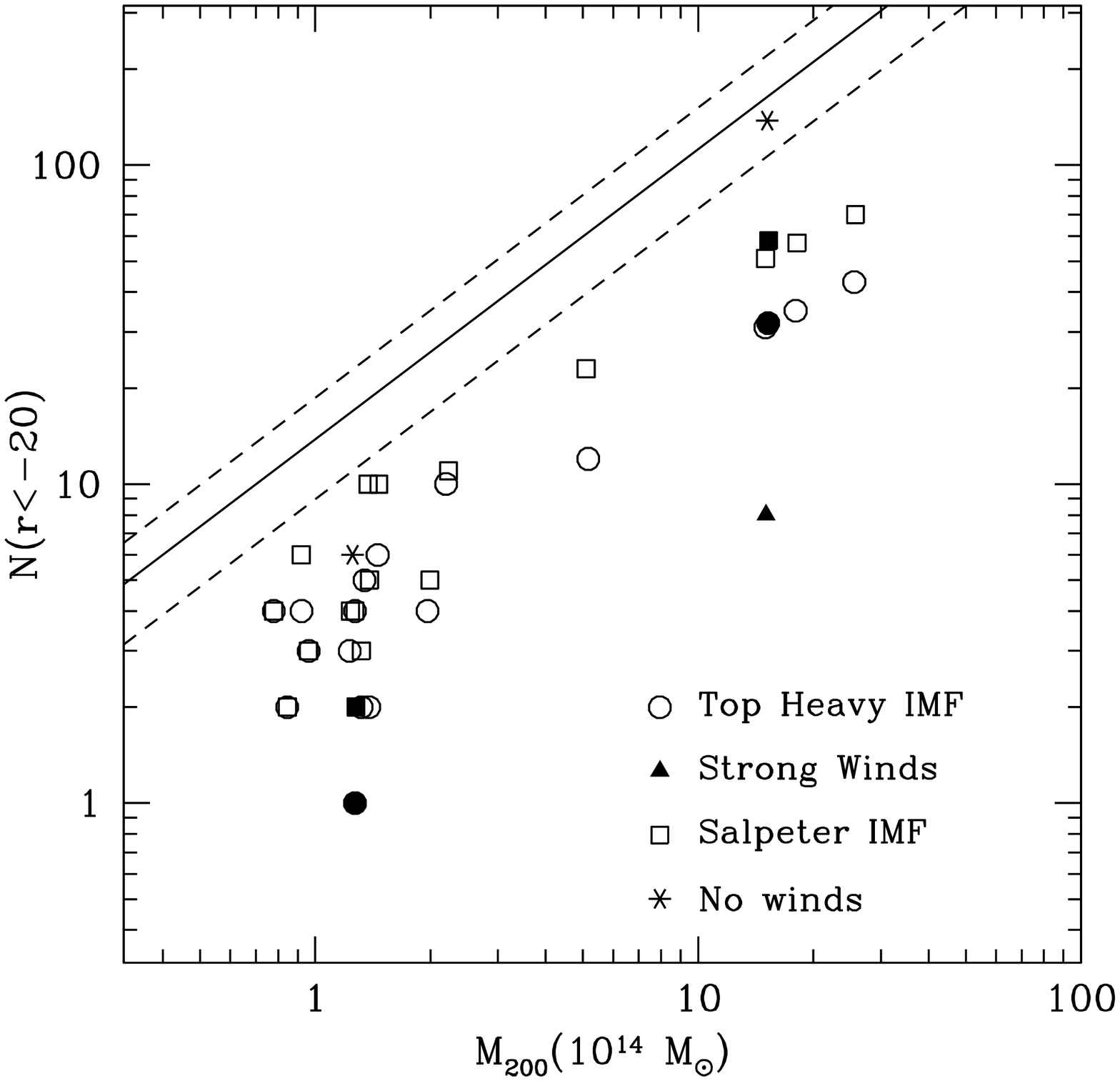,width=7.5cm} 
\psfig{file=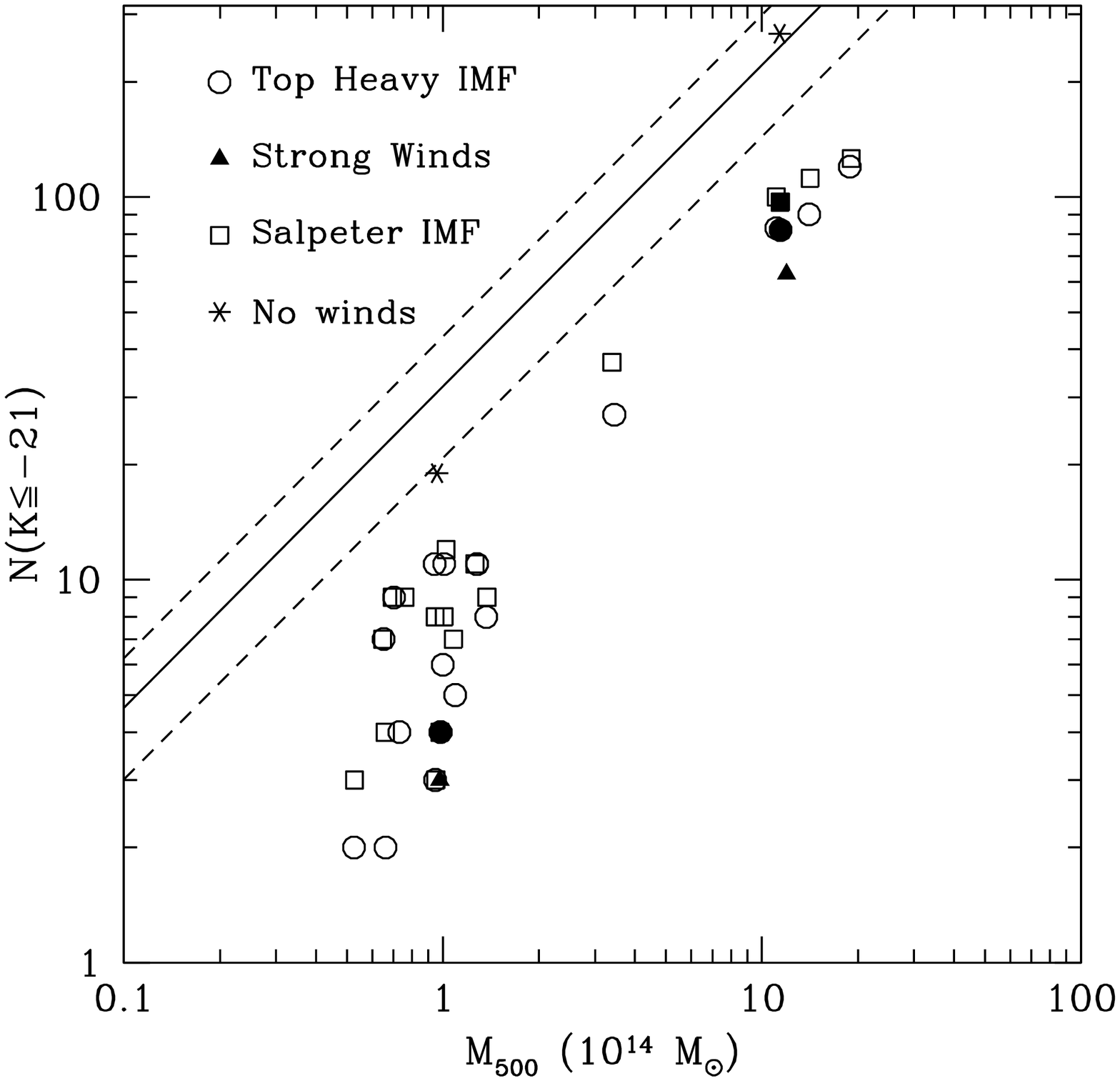,width=7.5cm} 
}}
\caption{The number of galaxies within clusters above a given
  luminosity limit. Left panel: results in the $r$ band, compared to
  the observational best--fitting result from SDSS data by
  \protect\cite{2006astro.ph..6260P} (the dashed lines mark the
  intrinsic scatter of the observational relation). Right panel:
  results in the $K$ band, compared to the observational best--fitting
  result from 2MASS data by
  \protect\cite{2004ApJ...610..745L}. Symbols for the simulations have
  the same meaning as in Figure \ref{fi:ml}.}
\label{fi:ngal} 
\end{figure*}

The reasonable level of agreement between the observed and the
simulated $M/L$ may suggest that our simulations produces a realistic
population of galaxies. However, as demonstrated in Figure
\ref{fi:ngal}, this is not the case. In this figure, we compare the
simulated and observed number of cluster galaxies, brighter than a
given luminosity limit, both in $i$ and in the $K$ bands. Clearly,
simulations underpredict such a number, by a factor $\sim 2$--3. This
result is at variance with respect to that from semi--analytical
models of galaxy formation, which instead predict the correct number
of cluster members
\citep[e.g.,][]{2004MNRAS.349.1101D,2005MNRAS.361..369L}. However,
semi--analytical models are generally successful in producing the
correct LF. They employ a suitable technique to track galaxies, 
based on the assumption that, once a ''satellite'' galaxy is formed
inside a DM halo, it preserves its identity and survive to a
possible disruption of the hosting halo \citep{SP01.2}. Accordingly,
the position of a galaxy is later assigned to the position of the DM
particle which was most bound within the DM halo before it was
disrupted, thereby preventing an excessive merging rate between
galaxies.

On the one hand, it is tempting to explain the lack of galaxies in our
simulations as the result of an excessive merging. On the other hand,
the inclusion of radiative cooling and star formation should produce
galaxies in our simulations which, in fact, survive to the merging of
DM halos, and behave as the ``satellite'' galaxies introduced in the
semi--analytical models. Clearly, a reason of concern in our
simulations is related to the force and mass resolutions adopted (see
Section \ref{sec:sims}), which may produce fragile galaxies and/or
induce spurious numerical overmerging. In the Appendix we present a
resolution study which is aimed at verifying whether and by how much
the cluster galaxy population changes when increasing the
resolution. After varying the mass resolution by a factor 45, and the
corresponding softening parameters by a factor $\simeq 3.6$, we find
no appreciable variations of the stellar mass function of cluster
galaxies. We also verified that the lack of galaxies is not related to
numerical heating induced by an non optimal choice of gravitational
softening \citep[e.g., ][]{1992MNRAS.257...11T}. After running a
series of simulations, using different choices for $\epsilon_{Pl}$,
we find that our softening choice is very close to that maximizing the
low end of the galaxy stellar mass function.

  As for the effect of feedback, a wind velocity of $500 \vel$ is
  large enough to devoid the gas content of galaxies with mass
  $M\mincir 10^{11}M_\odot$ and, therefore, to suppress the number of
  galaxies above the luminosity limits considered in
  Fig. \ref{fi:ngal}. Wind velocities this high are generally expected
  for starburst galaxies \citep[e.g.,][]{2003RMxAC..17...47H}, while
  they may be too high for the general galaxy population. In order to
  test this effect, we have performed simulations of g676 and g51 with
  Salpeter IMF in the extreme case in which winds are excluded. In
  these cases, the numbers of galaxies reported in Fig.\ref{fi:ngal}
  increase by more than a factor of two, thus bringing simulation
  results into much better agreement with observational data. However,
  the price to pay for this is the increased total luminosities,
  as a result of the larger number of stars formed, which introduces a
  tension between simulations and observations, as shown in
  Fig.\ref{fi:ml}.  The need to reconcile at the same time the number
  of galaxies and the total luminosity points toward a scenario in
  which feedback is relatively less effective in small galaxies, while
  being more effective in suppressing star formation in massive rare
  objects. Since massive galaxies are observed to be almost passively
  evolving, this implies that the required feedback mechanism should
  not be directly linked to star formation. In this respect, AGN have
  been suggested to be the natural source for this kind of feedback
  \citep[e.g.,][]{2006MNRAS.365...11C,2006MNRAS.370..645B}.

\subsection{The luminosity function}
\label{s:lf}
The luminosity function (LF hereafter) of cluster galaxies has been
the subject of numerous studies through the years \citep[e.g., ][,
and references
therein]{1978ApJ...223..765D,1989MNRAS.237..799C,1995A&A...297..610B,2002PASJ...54..515G,2003MNRAS.342..725D,2006A&A...445...29P}.
Despite this, a general consensus on a number of issues has still to
be reached.  Among them, we mention the LF universality among clusters
and between clusters and field, and the slope of the faint end. For
instance, \cite{2006A&A...445...29P} have recently analysed SDSS data
for a set of cluster selected in the X-ray band in the RASS. As a
result, they found that the LF is universal, once calculated within
the same physical radius, $r_{200}$ or $r_{500}$. Furthermore, the LF
can not be fitted by a single \cite{1976ApJ...203..297S} function,
since it displays a marked upturn at faint magnitudes. This result is
at variance with other analyses. For instance,
\cite{2000A&A...353..930A} performed a deep spectroscopic survey of
the Coma cluster and found no evidence for an upturn of the LF at
faint magnitudes.

\begin{figure*}
\centerline{
\hbox{
\psfig{file=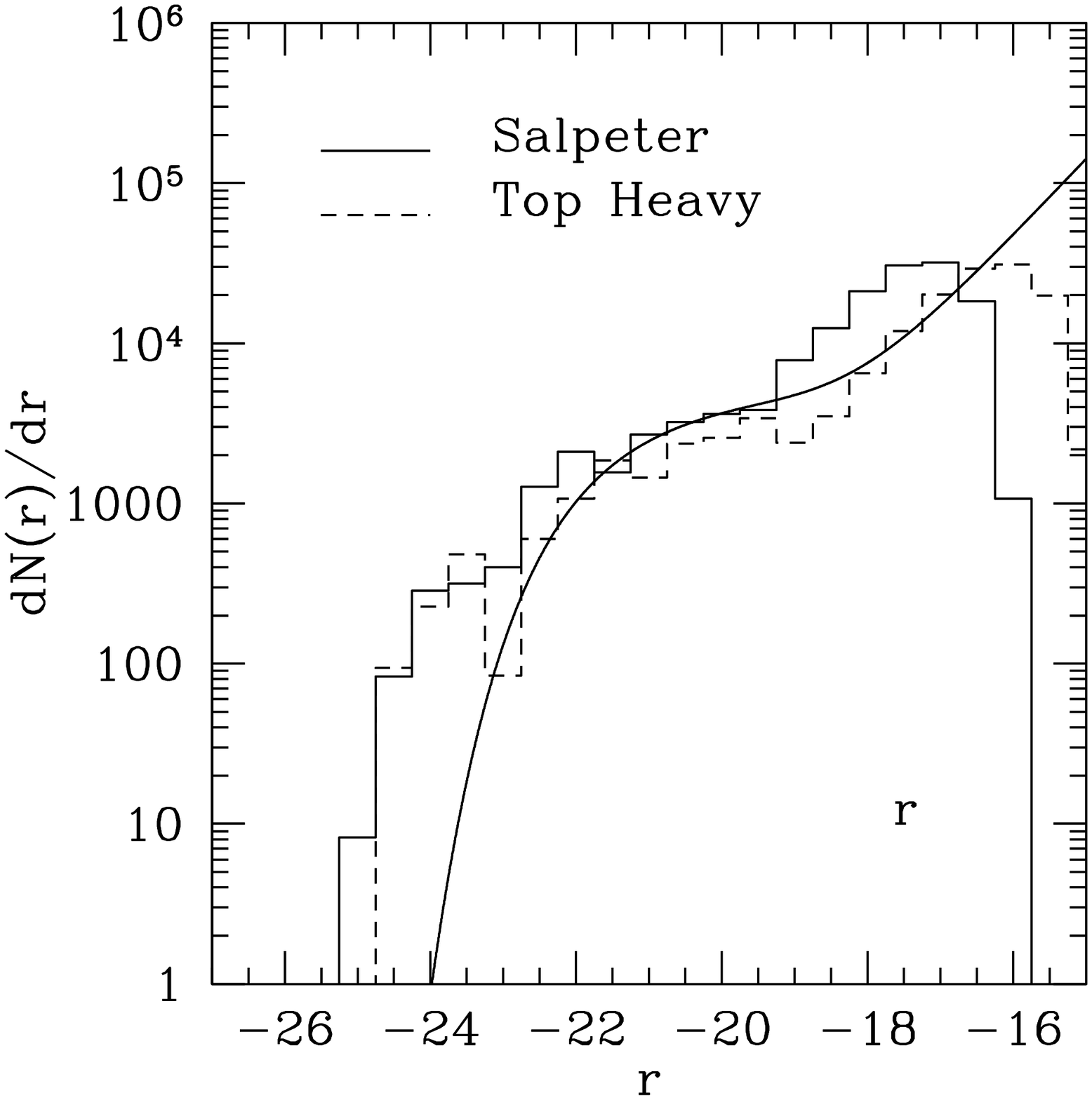,width=7.5cm} 
\psfig{file=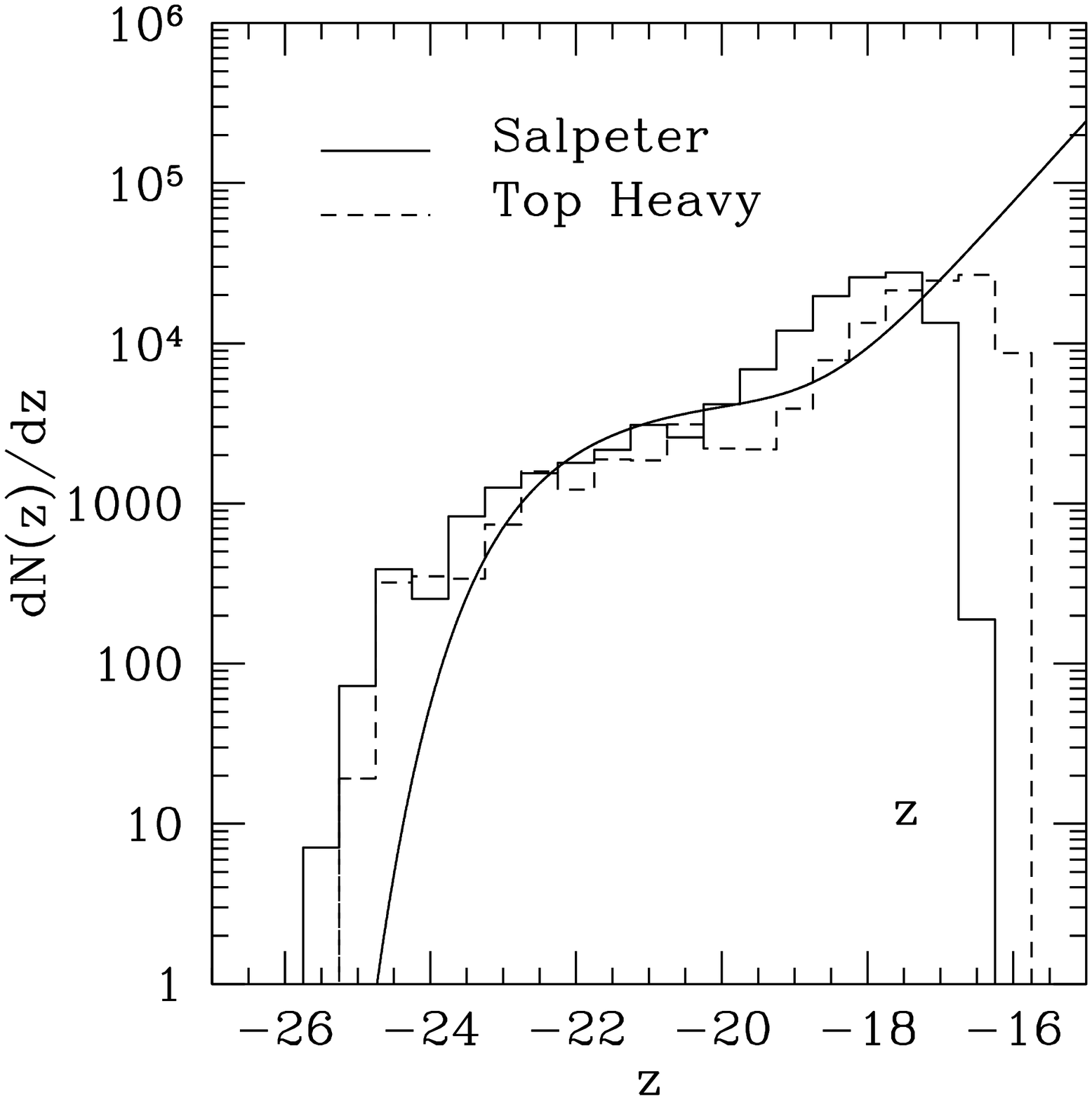,width=7.5cm} 
}}
\caption{The Comparison between the simulated (histograms) and the
  observed (curves) luminosity functions of cluster galaxies in the
  Sloan-$r$ (left panel) and $z$ (right panel) bands. The smooth
  curves are the best fit to the SDSS data analysed by
  \protect\cite{2006A&A...445...29P}. In each panel, the solid and the dashed
  histograms are for the Salpeter and for the top--heavy IMF,
  respectively. Consistent with the observational analysis, the
  brightest cluster galaxies (BCGs) are not included in the
  computation of the luminosity function.}
\label{fi:lf} 
\end{figure*}

In the following, we will discuss a comparison between the LF in our
simulated clusters and the observational results by
\cite{2006A&A...445...29P}.  To this purpose, we have computed the
simulated LF, within $r_{200}$, in the $r$ and $z$ bands, which are
two of the four SDSS bands where the analysis by
\cite{2006A&A...445...29P} has been performed. Consistently with their
approach, we have used the procedure introduced by
\cite{1989MNRAS.237..799C} to compute a composite luminosity function
from the contribution of clusters having different
richness. Accordingly, the number of galaxies $N_j$ within the $j$-th
luminosity bin is defined as
\be
N_j\,=\,{N_0\over m_j}\sum_i {N_{ij}\over N_{0,i}}\,.
\label{eq:clf}
\ee
Here $m_j$ is the number of clusters having galaxies in the $j$-th
luminosity bin, $N_{ij}$ is the number of galaxies in that luminosity
bin contributed by the $i$-th cluster, $N_{0,i}$ is the LF
normalization for the $i$-th cluster and $N_0=\sum_i
N_{0,i}$. Following \cite{2006A&A...445...29P}, we compute $N_{0,i}$
as the number of galaxies in the $i$-th cluster which are brighter
than $r,z=-19$. With this definition, each cluster is weighted
inversely to its richness, in such a way to avoid the richest clusters
to dominate the shape of the LF. Also, consistent with
\cite{2006A&A...445...29P}, we do not include the BCGs in the estimate
of the LF. Owing to the too small number of galaxies found in our
simulated clusters, we already know in advance that the normalization
of the simulated LF must be lower than the simulated one. Therefore,
we decide to normalize the simulated LF by hand, so that the LF for
the Salpeter IMF matches the observed one at $r=-20$ and $z=-20$. The
resulting rescaling factor is then used also to re-normalize the LF
for the runs with the top--heavy IMF. In this way, we preserve the
difference in normalization between the two series of runs, which is
induced by the different choices for the IMF.

\begin{figure}
\centerline{
\psfig{file=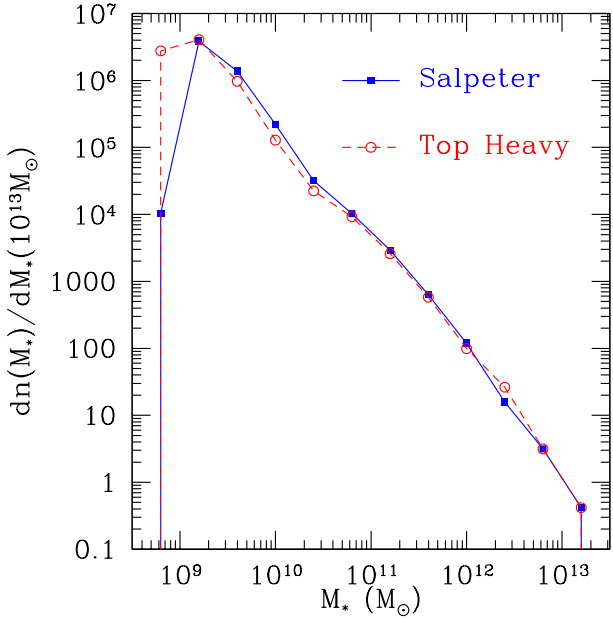,width=8.5cm}}
\caption{The combined stellar mass function for the galaxies
  identified within $r_{200}$ of all clusters. The solid and the
  dashed histograms correspond to the Salpeter and to the top-heavy
  IMF, respectively.}
\label{fi:massf} 
\end{figure}

The results of this comparison are shown in Figure \ref{fi:lf}. The
bright end of the simulated LF is clearly shallower than that of the
observed one. This is consistent with the picture that overcooling
takes place within the more massive halos, which hosts the brighter
galaxies. The simulated LF shows a steepening at the faint end, which
resembles that found by \cite{2006A&A...445...29P}. Looking at the
combined differential stellar mass function of all the cluster
galaxies (see Figure \ref{fi:massf}), we note an indication for
a steepening of its slope at the low-mass end, $M_*\mincir 3\times
10^{10}M_\odot$. It is this steepening which causes the corresponding
steepening of the luminosity functions. Quite interestingly, 
the faint end of the CMR (see Fig. \ref{fi:cmr_z0}) 
shows a population of small red galaxies, which are in fact associated
to the excess of faint galaxies shown by the luminosity function.  It
is tempting to make a correspondence between these galaxies and the
faint red galaxies which are claimed by \cite{2006A&A...445...29P} to
contribute to the steepening of their luminosity function. However, we
consider it as
premature to draw strong conclusions about the slope of the luminosity
function in simulations until the latter will be demonstrated to
roughly produce the correct total number of galaxies.

\begin{figure}
\centerline{
\psfig{file=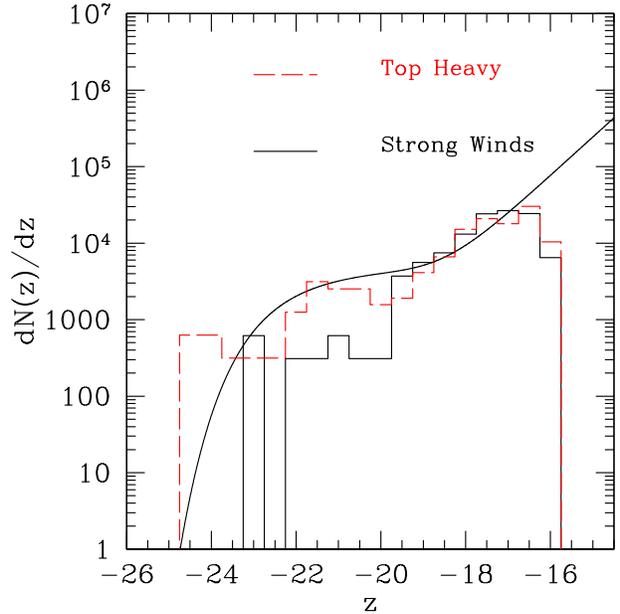,width=8.5cm} 
}
\caption{The effect of a stronger feedback on the $z$--band luminosity
  function. The histograms show the combined luminosity function for
  the g51 and g676 clusters in the case of standard feedback
  ($v_w=500\vel$, dashed line) and of strong feedback ($v_w=1000
  \vel$, solid line). The smooth curve is the best--fit to the SDSS
  data by \protect\cite{2006A&A...445...29P}.}
\label{fi:lf_sw} 
\end{figure}

A comparison between the LFs produced by the Salpeter and the
top--heavy IMF shows that the latter is shifted towards fainter
magnitude, especially in the faint end. This effect is also visible in
the corresponding stellar mass functions of the cluster galaxies (see
Fig.\ref{fi:massf}). While both IMFs produces indistinguishable mass
functions at the high end, galaxy masses for the top--heavy IMF tend
to have lower values. This difference is induced by the larger metal
content associated to the top--heavy IMF, which makes cooling more
efficient within halos near the resolution limit.
Finally, we show in Figure \ref{fi:lf_sw} the effect of increasing the
feedback efficiency on the LF. In this case, we use the same
normalization for the two IMFs, in order to directly see the effect of
changing the feedback strength. Quite interestingly, the effect is
that of suppressing the bright end of the LF, while leaving the faint
end almost unaffected. 

\subsection{Radial dependence of the galaxy population}
\label{s:rad}
A number of observations have established that the galaxy population
in clusters is characterized by the presence of color gradients, with
bluer galaxies preferentially avoiding to reside in the innermost
cluster regions \citep{1984ApJ...285..426B}. For instance,
\cite{2006MNRAS.366..645P} found a decreasing trend of the $B-R$ color
with cluster-centric distance for the galaxies lying on the CMR of
nearby optically selected clusters. Similar results have also been
found by \cite{1996ApJ...471..694A}, \cite{1997ApJ...478..462C} and
\cite{2005ApJ...627..186W} for moderately distant X--ray selected
clusters. Quite consistently, outer cluster regions are populated by a
larger fraction of blue galaxies \citep[e.g.,
][]{2004MNRAS.351..125D}, thus confirming that more external galaxies
are generally characterized by a relatively younger stellar
population. This effect may result both as a consequence of the
cluster environment, which excises star formation in infalling
galaxies, and/or due to an earlier formation epoch of galaxies
residing in the cluster center \citep[e.g.,
][]{2001ApJ...547..609E}. In general, the presence of a gradient in
the galaxy colors is naturally predicted by semi--analytical models of
galaxy formation \citep[e.g., ][]{2001MNRAS.323..999D}.

\begin{figure}
\centerline{
\psfig{file=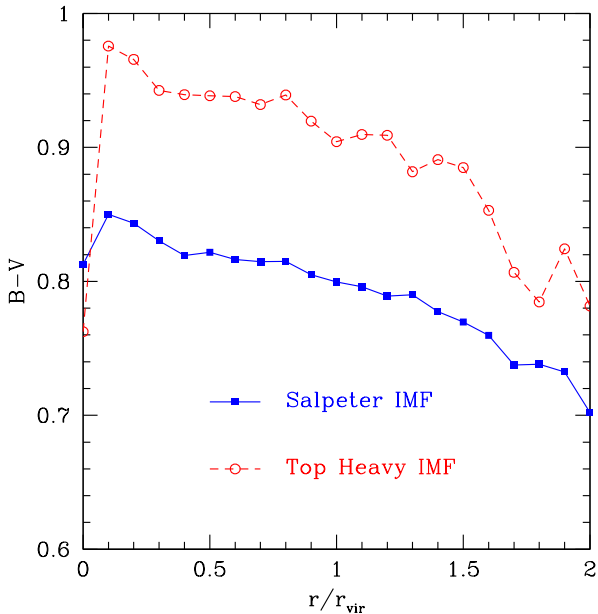,width=8.5cm} 
}
\caption{The radial dependence of galaxy colors, averaged over all
  simulated clusters. In each panel, solid lines with filled squares
  are for the Salpeter IMF, while dashed line with open circles are
  for the top--heavy IMF. The reported results are the average over
  all the simulated clusters.}
\label{fi:colrad} 
\end{figure}

In Figure \ref{fi:colrad} we show the radial variation of the $B-V$
color for all galaxies found in our set of simulated
clusters. Consistent with observational results, the mean galaxy
colors become bluer as we move towards the outer cluster
regions. Quite remarkably, this effect extends well beyond the virial
radius, thus implying that galaxies feel the cluster environment
already at fairly large distances. While the trend exists for both a
Salpeter and a top--heavy IMF, the latter generally predicts much
redder colors, consistent with the CMR results shown in
Fig. \ref{fi:cmr_z0}. Our results for the Salpeter IMF are consistent
with those reported by \cite{2001MNRAS.323..999D} for the
low--redshift bin ($0.18<z<0.3$) of the CNOC1 cluster sample.

We note a sudden inversion of the color gradients in the innermost
regions, where galaxies are characterized by much bluer colors. These
galaxies generally correspond to the cluster BCGs, which, as already
discussed are much bluer than expected from the CMR red sequence (see
Fig. \ref{fi:cmr_z0}). In fact, the galaxies identified in this region
correspond to the BCG, which, as we have already discussed, are
characterized by a strong excess of star formation. The effect is more
pronounced for the top--heavy IMF, which produces more metals and,
therefore, makes overcooling even stronger. This highlights once again
the presence of too high a cooling rate at the center of the simulated
clusters, which is not prevented by the model of SN feedback adopted
in our simulations.

\begin{figure}
\centerline{
\psfig{file=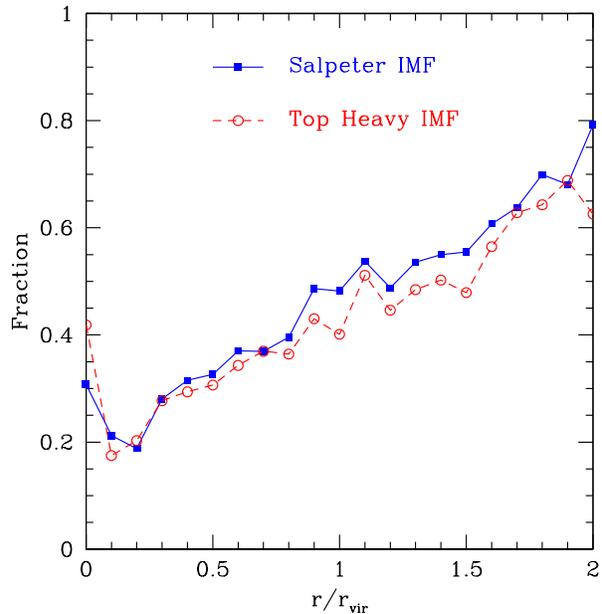,width=8.5cm} 
}
\caption{The fraction of galaxies younger that 8.5 Gyr, as a function
  of cluster--centric distance, in units of $r_{vir}$. Symbols and
  line types have the same meaning as in Figure \ref{fi:colrad}.}
\label{fi:agegrad} 
\end{figure}

The presence of color gradients corresponds to the presence of age
gradients. We show in Figure \ref{fi:agegrad} the radial dependence of
the fraction of galaxies younger than 8.5 Gyrs. Quite apparently,
there is a continuous trend for galaxies to be younger in the outer
cluster regions. The trend extends out to $2r_{vir}$, with no evidence
for convergence for a stable mean age in the field. This result
further confirms that the presence of a cluster induces environmental
effects in the galaxy population already at quite large
distances. Much like for the colors, we note an inversion of the trend
in the innermost regions, which is due to the excess star formation
taking place in the central BCGs. The fact that the inversion is more
pronounced for the top--heavy IMF is in line with its higher
enrichment, which makes gas cooling more efficient.

\begin{figure}
\centerline{
\psfig{file=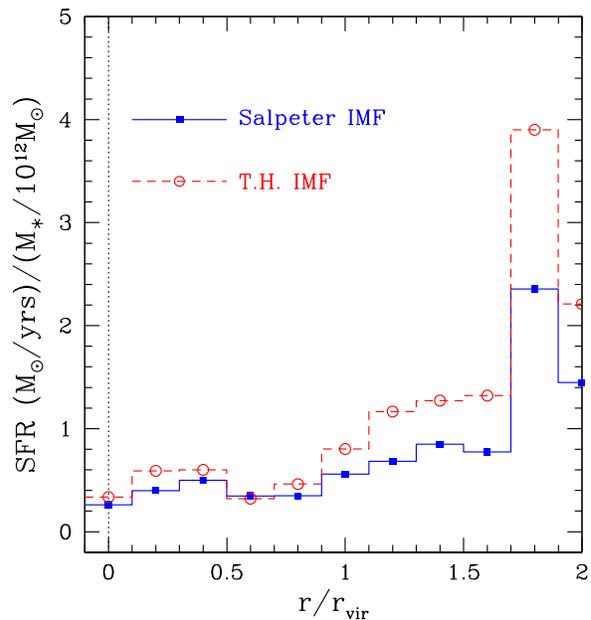,width=8.5cm} 
}
\caption{The specific mean star formation rate, averaged over all the
  simulated clusters, as a function of the cluster--centric
  distance. Symbols and line types have the same meaning as in Figure
  \ref{fi:colrad}. In the central bin we have excluded the
  contribution from the BCGs.}
\label{fi:sfr} 
\end{figure}

A consistent result also holds for the radial dependence of the star
formation rate (SFR). In Figure \ref{fi:sfr} we show the specific SFR
(i.e., the SFR per unit stellar mass) as a function of the
clustercentric distance. Once we exclude the contribution of the BCG
in the central bin, we observe a steady increase of the SFR toward
external cluster regions. In general, these results are in line with
observational evidences for a younger, more star forming galaxy
population in the cluster outskirts. For instance,
\cite{1997A&A...321...84B} analysed the galaxy population in the ESO
Nearby Abell Cluster Survey (ENACS) and found that emission--line
galaxies tends to underpopulate the central regions of clusters. 
\cite{1997ApJ...488L..75B} analysed data from the CNOC1 survey of
medium--distant galaxy clusters ($z\simeq 0.2$--0.6) and found
evidences for a continuous increase of the SFR out to $2r_{200}$. In a
similar way, \cite{2005ApJ...634..977M} analysed a large sample of
spectroscopic data, covering a $\sim 10$ Mpc regions around a Cl0024
at $z\simeq 0.4$. Again, they found that galaxies appear to be younger
at large radii. However, differently from our results, they detect
evidences for an increase of star formation around $r_{vir}$, possibly
triggered by the interaction with the dense environment of the ICM.

\section{Conclusions} \label{sec:conc}
\noindent
In this paper we have presented an analysis of the galaxy population
in cosmological hydrodynamical simulations of galaxy clusters at
$z=0$. The inclusion of a detailed treatment of stellar evolution and
chemical enrichment (\citealt{2004MNRAS.349L..19T}; Tornatore et al.,
in preparation) in the {\small GADGET-2} code \citep{2005MNRAS.364.1105S} has
allowed us to derive the properties of
galaxies in the optical/near--IR bands. In our simulations
each star particle is treated as a single stellar population (SSP)
characterized by a formation time and a metallicity. Based on this, we
apply the GALAXEV spectro-photometric code \citep{2003MNRAS.344.1000B}
to compute luminosities in different bands. Simulations have been
carried out for a representative set of galaxy clusters, containing 19
objects with mass $M_{200}$ ranging from $5\times 10^{13}\msun$ to
$1.8\times 10^{15}\msun$. All clusters have been simulated assuming
both a standard Salpeter IMF \citep{1955ApJ...121..161S} and a
power--law top--heavier IMF with exponent $x=0.95$
\citep{1987A&A...173...23A}. The main results of our analysis can be
summarized as follows.
\begin{description}
\item[1.] Both the color--magnitude relation (CMR) and the $M/L$ ratio
  are in reasonable agreement with observational data for a Salpeter
  IMF. In contrast, using a top--heavy IMF provides too high a
  metallicity of galaxies, which turns into too red colors. This
  spoils the agreement with the CMR and with the $M/L$ ratio in the
  $K$ band. The CMR is confirmed to be a metallicity sequence, in the
  sense that more enriched galaxies systematically populate the
  brighter redder part of the sequence.
\item[2.] Galaxies are systematically older and redder in central
  cluster regions, thus in keeping with observational results. This
  trend extends at least out to $2r_{vir}$, thus showing that the
  galaxy population feels the presence of a cluster well beyond its
  virial region. Due to the overcooling occurring in the central
  cluster regions, BCGs are always much bluer and more massive than
  observed, and characterized by too high a recent star formation. 
  Indeed, neglecting the contribution of stars formed at $z<1$
  produces significantly redder BCGs, thus in better agreement with
  observational data.
\item[3.] The number density profile of galaxies is confirmed to
  steepen with the galaxy stellar mass, approaching the DM profile for
  the galaxies with $M_*>10^{10}\msun$. However, when compared to the
  observations \citep[e.g.,][]{2006astro.ph..6260P}, the simulation profiles
  are flatter that the observed ones inside $0.4r_{200}$.
\item[4.] Simulated clusters have about three times fewer galaxies
  above a given luminosity limit than real clusters. We have verified
  that this disagreement is directly related neither to lack of mass
  and force resolution, nor to numerical gas heating due to a non
  optimal choice of the softening parameter. This leaves as further
  possibilities more subtle numerical effects or a better suited
  implementation of energy feedback.
\item[5.] The luminosity function (LF) is shallower than the observed
  one in the bright end, thus confirming that feedback is not strong
  enough to suppress cooling in the most massive halos. In the faint
  end, the LF steepens and indicates the presence of an excess of
  small red galaxies. Although this result resembles that found in
  observational data by \cite{2006A&A...445...29P}, we warn against
  its overinterpretation, in the light of the deficit of the overall
  number of galaxies found in the simulated clusters.
\end{description}

The results of our analysis support the capability of hydrodynamical
simulations of galaxy clusters to reproduce the general trends
characterizing their galaxy population. Therefore, simulations in
which the properties of the stellar population are self--consistently
predicted from the gas dynamics, can provide an alternative and
complementary approach to semi--analytical methods. However, at
present, our simulations have two main limitations in accounting for
the observed properties of the cluster galaxy population.

First of all, the deficit of galaxies produced in our
simulations suggests that they are not able to produce galaxies
which are resistant enough to survive the tidal field of the cluster
environment. Indeed, the shallow galaxy number density profile in
central cluster regions shows that the problem is more apparent
where effects of galaxy disruption are expected to be stronger. 
This result on the lack of galaxies is at variance with respect to the
predictions of semi--analytical models (SAM) of galaxy formation,
which produce roughly the correct number of cluster galaxies
\citep[e.g., ][]{2004MNRAS.349.1101D,2005MNRAS.361..369L}.
Comparisons between the galaxy populations predicted by hydrodynamical
simulations and by SAM have shown a reasonable level of
agreement. However, these comparisons have been always performed by
excluding the effect of energy feedback \citep[e.g.,
][]{2003MNRAS.338..913H,2006astro.ph..5750C} or explicit conversion of
cooled gas particles into collisionless stars \citep[e.g.,
][]{YO02.1}.

Furthermore, we always find that the brightest cluster galaxies (BCGs)
are much bluer and star forming that observed. While the adopted
feedback scheme, based on galactic winds, is efficient in regulating
star formation for the bulk of the galaxy population, it is not able
to quench low--redshift star formation in the central cluster regions,
to a level consistent with observations. Clearly, the required
feedback mechanism should be such to leave the bulk of the galaxy
population unaffected while acting only on the very high end of the
galaxy mass distribution. Feedback from central AGN represents the
natural solution and provides in principle a large enough energy
budget \citep[e.g., ][]{2006astro.ph..5323R}. Its detailed
implementation in cosmological hydrodynamical simulations require
understanding in detail the mechanisms for the thermalization of this
energy in the diffuse medium \citep[e.g.,
][]{2005A&A...429..399Z,2006MNRAS.366..397S}.

An ambitious goal for hydrodynamical simulations of the next
generation will be that of describing in detail the complex interplay
between the history of star formation and the thermodynamical and
chemical evolution of the diffuse cosmic baryons. For this to be
reached, it is mandatory that simulations are able to produce a
realistic population of galaxies, both in terms of color and of
luminosity and mass distribution. Therefore, simulation codes will be
required on the one hand to have numerical effects under exquisite
control and, on the other hand, to include physically motivated
schemes of energy and metal feedback.

\section*{acknowledgments}
The simulations were carried out at the ``Centro Interuniversitario
del Nord-Est per il Calcolo Elettronico'' (CINECA, Bologna), with CPU
time assigned under INAF/CINECA and University-of-Trieste/CINECA
grants. We are grateful to Volker Springel, who provided us with the
non public version of the {\small GADGET-2} code. We acknowledge
useful discussions with Gabriella De Lucia, Antonaldo Diaferio, Marisa
Girardi, Andrey Kravtsov, Bianca Poggianti and Paola Popesso. This
work has been partially supported by the PD-51 INFN grant.

\bibliographystyle{mn2e}
\bibliography{master}

\begin{thebibliography}{}

\bibitem[\protect\citeauthoryear{{Abadi}, {Moore} \& {Bower}}{{Abadi}
  et~al.}{1999}]{1999MNRAS.308..947A}
{Abadi} M.~G.,  {Moore} B.,    {Bower} R.~G.,  1999, \mnras, 308, 947

\bibitem[\protect\citeauthoryear{{Abraham}, {Smecker-Hane}, {Hutchings},
  {Carlberg}, {Yee}, {Ellingson}, {Morris}, {Oke} \& {Rigler}}{{Abraham}
  et~al.}{1996}]{1996ApJ...471..694A}
{Abraham} R.~G.,  {Smecker-Hane} T.~A.,  {Hutchings} J.~B.,  {Carlberg} R.~G.,
  {Yee} H.~K.~C.,  {Ellingson} E.,  {Morris} S.,  {Oke} J.~B.,    {Rigler} M.,
  1996, \apj, 471, 694

\bibitem[\protect\citeauthoryear{{Adami}, {Mazure}, {Biviano}, {Katgert} \&
  {Rhee}}{{Adami} et~al.}{1998}]{1998A&A...331..493A}
{Adami} C.,  {Mazure} A.,  {Biviano} A.,  {Katgert} P.,    {Rhee} G.,  1998,
  \aap, 331, 493

\bibitem[\protect\citeauthoryear{{Adami}, {Ulmer}, {Durret}, {Nichol},
  {Mazure}, {Holden}, {Romer} \& {Savine}}{{Adami}
  et~al.}{2000}]{2000A&A...353..930A}
{Adami} C.,  {Ulmer} M.~P.,  {Durret} F.,  {Nichol} R.~C.,  {Mazure} A.,
  {Holden} B.~P.,  {Romer} A.~K.,    {Savine} C.,  2000, \aap, 353, 930

\bibitem[\protect\citeauthoryear{{Aguirre}, {Hernquist}, {Schaye}, {Weinberg},
  {Katz} \& {Gardner}}{{Aguirre} et~al.}{2001}]{2001ApJ...560..599A}
{Aguirre} A.,  {Hernquist} L.,  {Schaye} J.,  {Weinberg} D.~H.,  {Katz} N.,
  {Gardner} J.,  2001, \apj, 560, 599

\bibitem[\protect\citeauthoryear{{Andreon}}{{Andreon}}{2003}]{2003A&A...409...%
37A}
{Andreon} S.,  2003, \aap, 409, 37

\bibitem[\protect\citeauthoryear{{Andreon}, {Willis}, {Quintana}, {Valtchanov},
  {Pierre} \& {Pacaud}}{{Andreon} et~al.}{2004}]{2004MNRAS.353..353A}
{Andreon} S.,  {Willis} J.,  {Quintana} H.,  {Valtchanov} I.,  {Pierre} M.,
  {Pacaud} F.,  2004, \mnras, 353, 353

\bibitem[\protect\citeauthoryear{{Arimoto} \& {Yoshii}}{{Arimoto} \&
  {Yoshii}}{1987}]{1987A&A...173...23A}
{Arimoto} N.,  {Yoshii} Y.,  1987, \aap, 173, 23

\bibitem[\protect\citeauthoryear{{Bahcall} \& {Comerford}}{{Bahcall} \&
  {Comerford}}{2002}]{2002ApJ...565L...5B}
{Bahcall} N.~A.,  {Comerford} J.~M.,  2002, \apjl, 565, L5

\bibitem[\protect\citeauthoryear{{Balogh}, {Morris}, {Yee}, {Carlberg} \&
  {Ellingson}}{{Balogh} et~al.}{1997}]{1997ApJ...488L..75B}
{Balogh} M.~L.,  {Morris} S.~L.,  {Yee} H.~K.~C.,  {Carlberg} R.~G.,
  {Ellingson} E.,  1997, \apjl, 488, L75+

\bibitem[\protect\citeauthoryear{{Barrientos}, {Schade}, {L{\'o}pez-Cruz} \&
  {Quintana}}{{Barrientos} et~al.}{2004}]{2004ApJS..153..397B}
{Barrientos} L.~F.,  {Schade} D.,  {L{\'o}pez-Cruz} O.,    {Quintana} H.,
  2004, \apjs, 153, 397

\bibitem[\protect\citeauthoryear{{Biviano}, {Durret}, {Gerbal}, {Le Fevre},
  {Lobo}, {Mazure} \& {Slezak}}{{Biviano} et~al.}{1995}]{1995A&A...297..610B}
{Biviano} A.,  {Durret} F.,  {Gerbal} D.,  {Le Fevre} O.,  {Lobo} C.,  {Mazure}
  A.,    {Slezak} E.,  1995, \aap, 297, 610

\bibitem[\protect\citeauthoryear{{Biviano}, {Katgert}, {Mazure}, {Moles}, {den
  Hartog}, {Perea} \& {Focardi}}{{Biviano} et~al.}{1997}]{1997A&A...321...84B}
{Biviano} A.,  {Katgert} P.,  {Mazure} A.,  {Moles} M.,  {den Hartog} R.,
  {Perea} J.,    {Focardi} P.,  1997, \aap, 321, 84

\bibitem[\protect\citeauthoryear{{Biviano}, {Murante}, {Borgani}, {Diaferio},
  {Dolag} \& {Girardi}}{{Biviano} et~al.}{2006}]{2006astro.ph..5151B}
{Biviano} A.,  {Murante} G.,  {Borgani} S.,  {Diaferio} A.,  {Dolag} K.,
  {Girardi} M.,  2006, ArXiv Astrophysics e-prints

\bibitem[\protect\citeauthoryear{{Blakeslee}, {Franx}, {Postman} \& {et
  al.}}{{Blakeslee} et~al.}{2003}]{2003ApJ...596L.143B}
{Blakeslee} J.~P.,  {Franx} M.,  {Postman} M.,    {et al.} 2003, \apjl, 596,
  L143

\bibitem[\protect\citeauthoryear{{Borgani}, {Dolag}, {Murante}, {Cheng},
  {Springel}, {Diaferio}, {Moscardini}, {Tormen}, {Tornatore} \&
  {Tozzi}}{{Borgani} et~al.}{2006}]{2006MNRAS.367.1641B}
{Borgani} S.,  {Dolag} K.,  {Murante} G.,  {Cheng} L.-M.,  {Springel} V.,
  {Diaferio} A.,  {Moscardini} L.,  {Tormen} G.,  {Tornatore} L.,    {Tozzi}
  P.,  2006, \mnras, 367, 1641

\bibitem[\protect\citeauthoryear{{Bower}, {Benson}, {Malbon}, {Helly}, {Frenk},
  {Baugh}, {Cole} \& {Lacey}}{{Bower} et~al.}{2006}]{2006MNRAS.370..645B}
{Bower} R.~G.,  {Benson} A.~J.,  {Malbon} R.,  {Helly} J.~C.,  {Frenk} C.~S.,
  {Baugh} C.~M.,  {Cole} S.,    {Lacey} C.~G.,  2006, \mnras, 370, 645

\bibitem[\protect\citeauthoryear{{Bower}, {Lucey} \& {Ellis}}{{Bower}
  et~al.}{1992}]{1992MNRAS.254..601B}
{Bower} R.~G.,  {Lucey} J.~R.,    {Ellis} R.~S.,  1992, \mnras, 254, 601

\bibitem[\protect\citeauthoryear{{Bregman}, {Fabian}, {Miller} \&
  {Irwin}}{{Bregman} et~al.}{2006}]{2006astro.ph..2323B}
{Bregman} J.~N.,  {Fabian} A.~C.,  {Miller} E.~D.,    {Irwin} J.~A.,  2006,
  ArXiv Astrophysics e-prints

\bibitem[\protect\citeauthoryear{{Bruzual} \& {Charlot}}{{Bruzual} \&
  {Charlot}}{2003}]{2003MNRAS.344.1000B}
{Bruzual} G.,  {Charlot} S.,  2003, \mnras, 344, 1000

\bibitem[\protect\citeauthoryear{{Butcher} \& {Oemler}}{{Butcher} \&
  {Oemler}}{1978}]{1978ApJ...226..559B}
{Butcher} H.,  {Oemler} A.,  1978, \apj, 226, 559

\bibitem[\protect\citeauthoryear{{Butcher} \& {Oemler}}{{Butcher} \&
  {Oemler}}{1984}]{1984ApJ...285..426B}
{Butcher} H.,  {Oemler} A.,  1984, \apj, 285, 426

\bibitem[\protect\citeauthoryear{{Carlberg}, {Yee} \& {Ellingson}}{{Carlberg}
  et~al.}{1997}]{1997ApJ...478..462C}
{Carlberg} R.~G.,  {Yee} H.~K.~C.,    {Ellingson} E.,  1997, \apj, 478, 462

\bibitem[\protect\citeauthoryear{{Casagrande} \& {Diaferio}}{{Casagrande} \&
  {Diaferio}}{2006}]{2006astro.ph..6591C}
{Casagrande} L.,  {Diaferio} A.,  2006, ArXiv Astrophysics e-prints

\bibitem[\protect\citeauthoryear{{Cattaneo}, {Blaizot}, {Weinberg}, {Colombi},
  {Dave}, {Devriendt}, {Guiderdoni}, {Katz} \& {Keres}}{{Cattaneo}
  et~al.}{2006}]{2006astro.ph..5750C}
{Cattaneo} A.,  {Blaizot} J.,  {Weinberg} D.~H.,  {Colombi} S.,  {Dave} R.,
  {Devriendt} J.,  {Guiderdoni} B.,  {Katz} N.,    {Keres} D.,  2006, ArXiv
  Astrophysics e-prints

\bibitem[\protect\citeauthoryear{{Chiappini}, {Matteucci} \&
  {Gratton}}{{Chiappini} et~al.}{1997}]{1997ApJ...477..765C}
{Chiappini} C.,  {Matteucci} F.,    {Gratton} R.,  1997, \apj, 477, 765

\bibitem[\protect\citeauthoryear{{Cole}, {Lacey}, {Baugh} \& {Frenk}}{{Cole}
  et~al.}{2000}]{2000MNRAS.319..168C}
{Cole} S.,  {Lacey} C.~G.,  {Baugh} C.~M.,    {Frenk} C.~S.,  2000, \mnras,
  319, 168

\bibitem[\protect\citeauthoryear{{Colless}}{{Colless}}{1989}]{1989MNRAS.237..7%
99C}
{Colless} M.,  1989, \mnras, 237, 799

\bibitem[\protect\citeauthoryear{{Cora}}{{Cora}}{2006}]{2006MNRAS.tmp..453C}
{Cora} S.~A.,  2006, \mnras, pp 453--+

\bibitem[\protect\citeauthoryear{{Croton}, {Springel}, {White}, {De Lucia},
  {Frenk}, {Gao}, {Jenkins}, {Kauffmann}, {Navarro} \& {Yoshida}}{{Croton}
  et~al.}{2006}]{2006MNRAS.365...11C}
{Croton} D.~J.,  {Springel} V.,  {White} S.~D.~M.,  {De Lucia} G.,  {Frenk}
  C.~S.,  {Gao} L.,  {Jenkins} A.,  {Kauffmann} G.,  {Navarro} J.~F.,
  {Yoshida} N.,  2006, \mnras, 365, 11

\bibitem[\protect\citeauthoryear{{De Lucia}, {Kauffmann}, {Springel}, {White},
  {Lanzoni}, {Stoehr}, {Tormen} \& {Yoshida}}{{De Lucia}
  et~al.}{2004}]{2004MNRAS.348..333D}
{De Lucia} G.,  {Kauffmann} G.,  {Springel} V.,  {White} S.~D.~M.,  {Lanzoni}
  B.,  {Stoehr} F.,  {Tormen} G.,    {Yoshida} N.,  2004, \mnras, 348, 333

\bibitem[\protect\citeauthoryear{{De Lucia}, {Kauffmann} \& {White}}{{De Lucia}
  et~al.}{2004}]{2004MNRAS.349.1101D}
{De Lucia} G.,  {Kauffmann} G.,    {White} S.~D.~M.,  2004, \mnras, 349, 1101

\bibitem[\protect\citeauthoryear{{De Propris}, {Colless}, {Driver} \& {et
  al.}}{{De Propris} et~al.}{2003}]{2003MNRAS.342..725D}
{De Propris} R.,  {Colless} M.,  {Driver} S.~P.,    {et al.} 2003, \mnras, 342,
  725

\bibitem[\protect\citeauthoryear{{De Propris}, {Colless}, {Peacock} \& {et
  al.}}{{De Propris} et~al.}{2004}]{2004MNRAS.351..125D}
{De Propris} R.,  {Colless} M.,  {Peacock} J.~A.,    {et al.} 2004, \mnras,
  351, 125

\bibitem[\protect\citeauthoryear{{Diaferio}, {Kauffmann}, {Balogh}, {White},
  {Schade} \& {Ellingson}}{{Diaferio} et~al.}{2001}]{2001MNRAS.323..999D}
{Diaferio} A.,  {Kauffmann} G.,  {Balogh} M.~L.,  {White} S.~D.~M.,  {Schade}
  D.,    {Ellingson} E.,  2001, \mnras, 323, 999

\bibitem[\protect\citeauthoryear{{Diemand}, {Moore} \& {Stadel}}{{Diemand}
  et~al.}{2004}]{2004MNRAS.352..535D}
{Diemand} J.,  {Moore} B.,    {Stadel} J.,  2004, \mnras, 352, 535

\bibitem[\protect\citeauthoryear{{Dolag}, {Vazza}, {Brunetti} \&
  {Tormen}}{{Dolag} et~al.}{2005}]{2005MNRAS.364..753D}
{Dolag} K.,  {Vazza} F.,  {Brunetti} G.,    {Tormen} G.,  2005, \mnras, 364,
  753

\bibitem[\protect\citeauthoryear{{Domainko}, {Mair}, {Kapferer}, {van Kampen},
  {Kronberger}, {Schindler}, {Kimeswenger}, {Ruffert} \& {Mangete}}{{Domainko}
  et~al.}{2005}]{2005astro.ph..7605D}
{Domainko} W.,  {Mair} M.,  {Kapferer} W.,  {van Kampen} E.,  {Kronberger} T.,
  {Schindler} S.,  {Kimeswenger} S.,  {Ruffert} M.,    {Mangete} O.~E.,  2005,
  ArXiv Astrophysics e-prints

\bibitem[\protect\citeauthoryear{{Dressler}}{{Dressler}}{1978}]{1978ApJ...223.%
.765D}
{Dressler} A.,  1978, \apj, 223, 765

\bibitem[\protect\citeauthoryear{{Dressler}}{{Dressler}}{1980}]{1980ApJ...236.%
.351D}
{Dressler} A.,  1980, \apj, 236, 351

\bibitem[\protect\citeauthoryear{{Driver}}{{Driver}}{2004}]{2004PASA...21..344%
D}
{Driver} S.,  2004, Publications of the Astronomical Society of Australia, 21,
  344

\bibitem[\protect\citeauthoryear{{Ellingson}, {Lin}, {Yee} \&
  {Carlberg}}{{Ellingson} et~al.}{2001}]{2001ApJ...547..609E}
{Ellingson} E.,  {Lin} H.,  {Yee} H.~K.~C.,    {Carlberg} R.~G.,  2001, \apj,
  547, 609

\bibitem[\protect\citeauthoryear{{Faltenbacher}, {Kravtsov}, {Nagai} \&
  {Gottl{\"o}ber}}{{Faltenbacher} et~al.}{2005}]{2005MNRAS.358..139F}
{Faltenbacher} A.,  {Kravtsov} A.~V.,  {Nagai} D.,    {Gottl{\"o}ber} S.,
  2005, \mnras, 358, 139

\bibitem[\protect\citeauthoryear{{Frenk}, {Evrard}, {White} \&
  {Summers}}{{Frenk} et~al.}{1996}]{1996ApJ...472..460F}
{Frenk} C.~S.,  {Evrard} A.~E.,  {White} S.~D.~M.,    {Summers} F.~J.,  1996,
  \apj, 472, 460

\bibitem[\protect\citeauthoryear{{Gao}, {De Lucia}, {White} \& {Jenkins}}{{Gao}
  et~al.}{2004}]{2004MNRAS.352L...1G}
{Gao} L.,  {De Lucia} G.,  {White} S.~D.~M.,    {Jenkins} A.,  2004, \mnras,
  352, L1

\bibitem[\protect\citeauthoryear{{Ghigna}, {Moore}, {Governato}, {Lake},
  {Quinn} \& {Stadel}}{{Ghigna} et~al.}{2000}]{2000ApJ...544..616G}
{Ghigna} S.,  {Moore} B.,  {Governato} F.,  {Lake} G.,  {Quinn} T.,    {Stadel}
  J.,  2000, \apj, 544, 616

\bibitem[\protect\citeauthoryear{{Girardi}, {Borgani}, {Giuricin},
  {Mardirossian} \& {Mezzetti}}{{Girardi} et~al.}{2000}]{2000ApJ...530...62G}
{Girardi} M.,  {Borgani} S.,  {Giuricin} G.,  {Mardirossian} F.,    {Mezzetti}
  M.,  2000, \apj, 530, 62

\bibitem[\protect\citeauthoryear{{Girardi}, {Manzato}, {Mezzetti}, {Giuricin}
  \& {Limboz}}{{Girardi} et~al.}{2002}]{2002ApJ...569..720G}
{Girardi} M.,  {Manzato} P.,  {Mezzetti} M.,  {Giuricin} G.,    {Limboz} F.,
  2002, \apj, 569, 720

\bibitem[\protect\citeauthoryear{{Gladders} \& {Yee}}{{Gladders} \&
  {Yee}}{2005}]{2005ApJS..157....1G}
{Gladders} M.~D.,  {Yee} H.~K.~C.,  2005, \apjs, 157, 1

\bibitem[\protect\citeauthoryear{{Goto}, {Okamura}, {McKay}, {Bahcall},
  {Annis}, {Bernard}, {Brinkmann}, {G{\'o}mez}, {Hansen}, {Kim}, {Sekiguchi} \&
  {Sheth}}{{Goto} et~al.}{2002}]{2002PASJ...54..515G}
{Goto} T.,  {Okamura} S.,  {McKay} T.~A.,  {Bahcall} N.~A.,  {Annis} J.,
  {Bernard} M.,  {Brinkmann} J.,  {G{\'o}mez} P.~L.,  {Hansen} S.,  {Kim}
  R.~S.~J.,  {Sekiguchi} M.,    {Sheth} R.~K.,  2002, \pasj, 54, 515

\bibitem[\protect\citeauthoryear{{Greggio} \& {Renzini}}{{Greggio} \&
  {Renzini}}{1983}]{1983A&A...118..217G}
{Greggio} L.,  {Renzini} A.,  1983, \aap, 118, 217

\bibitem[\protect\citeauthoryear{{Gunn} \& {Gott}}{{Gunn} \&
  {Gott}}{1972}]{1972ApJ...176....1G}
{Gunn} J.~E.,  {Gott} J.~R.~I.,  1972, \apj, 176, 1

\bibitem[\protect\citeauthoryear{{Haardt} \& {Madau}}{{Haardt} \&
  {Madau}}{1996}]{1996ApJ...461...20H}
{Haardt} F.,  {Madau} P.,  1996, \apj, 461, 20

\bibitem[\protect\citeauthoryear{{Heckman}}{{Heckman}}{2003}]{2003RMxAC..17...%
47H}
{Heckman} T.~M.,  2003, in {Avila-Reese} V.,  {Firmani} C.,  {Frenk} C.~S.,
  {Allen} C.,  eds, Revista Mexicana de Astronomia y Astrofisica Conference
  Series {Starburst-Driven Galactic Winds}.
pp 47--55

\bibitem[\protect\citeauthoryear{{Heger} \& {Woosley}}{{Heger} \&
  {Woosley}}{2002}]{2002ApJ...567..532H}
{Heger} A.,  {Woosley} S.~E.,  2002, \apj, 567, 532

\bibitem[\protect\citeauthoryear{{Helly}, {Cole}, {Frenk}, {Baugh}, {Benson},
  {Lacey} \& {Pearce}}{{Helly} et~al.}{2003}]{2003MNRAS.338..913H}
{Helly} J.~C.,  {Cole} S.,  {Frenk} C.~S.,  {Baugh} C.~M.,  {Benson} A.,
  {Lacey} C.,    {Pearce} F.~R.,  2003, \mnras, 338, 913

\bibitem[\protect\citeauthoryear{{Johnstone}, {Fabian} \& {Nulsen}}{{Johnstone}
  et~al.}{1987}]{1987MNRAS.224...75J}
{Johnstone} R.~M.,  {Fabian} A.~C.,    {Nulsen} P.~E.~J.,  1987, \mnras, 224,
  75

\bibitem[\protect\citeauthoryear{{Kauffmann}, {Colberg}, {Diaferio} \&
  {White}}{{Kauffmann} et~al.}{1999}]{1999MNRAS.303..188K}
{Kauffmann} G.,  {Colberg} J.~M.,  {Diaferio} A.,    {White} S.~D.~M.,  1999,
  \mnras, 303, 188

\bibitem[\protect\citeauthoryear{{Kauffmann}, {White} \&
  {Guiderdoni}}{{Kauffmann} et~al.}{1993}]{1993MNRAS.264..201K}
{Kauffmann} G.,  {White} S.~D.~M.,    {Guiderdoni} B.,  1993, \mnras, 264, 201

\bibitem[\protect\citeauthoryear{{Kenney}, {van Gorkom} \& {Vollmer}}{{Kenney}
  et~al.}{2004}]{2004AJ....127.3361K}
{Kenney} J.~D.~P.,  {van Gorkom} J.~H.,    {Vollmer} B.,  2004, \aj, 127, 3361

\bibitem[\protect\citeauthoryear{{Kodama} \& {Arimoto}}{{Kodama} \&
  {Arimoto}}{1997}]{1997A&A...320...41K}
{Kodama} T.,  {Arimoto} N.,  1997, \aap, 320, 41

\bibitem[\protect\citeauthoryear{{Kravtsov}, {Berlind}, {Wechsler}, {Klypin},
  {Gottl{\"o}ber}, {Allgood} \& {Primack}}{{Kravtsov}
  et~al.}{2004}]{2004ApJ...609...35K}
{Kravtsov} A.~V.,  {Berlind} A.~A.,  {Wechsler} R.~H.,  {Klypin} A.~A.,
  {Gottl{\"o}ber} S.,  {Allgood} B.,    {Primack} J.~R.,  2004, \apj, 609, 35

\bibitem[\protect\citeauthoryear{{Lanzoni}, {Guiderdoni}, {Mamon}, {Devriendt}
  \& {Hatton}}{{Lanzoni} et~al.}{2005}]{2005MNRAS.361..369L}
{Lanzoni} B.,  {Guiderdoni} B.,  {Mamon} G.~A.,  {Devriendt} J.,    {Hatton}
  S.,  2005, \mnras, 361, 369

\bibitem[\protect\citeauthoryear{{Lin}, {Mohr} \& {Stanford}}{{Lin}
  et~al.}{2003}]{2003ApJ...591..749L}
{Lin} Y.-T.,  {Mohr} J.~J.,    {Stanford} S.~A.,  2003, \apj, 591, 749

\bibitem[\protect\citeauthoryear{{Lin}, {Mohr} \& {Stanford}}{{Lin}
  et~al.}{2004}]{2004ApJ...610..745L}
{Lin} Y.-T.,  {Mohr} J.~J.,    {Stanford} S.~A.,  2004, \apj, 610, 745

\bibitem[\protect\citeauthoryear{{L{\'o}pez-Cruz}, {Barkhouse} \&
  {Yee}}{{L{\'o}pez-Cruz} et~al.}{2004}]{2004ApJ...614..679L}
{L{\'o}pez-Cruz} O.,  {Barkhouse} W.~A.,    {Yee} H.~K.~C.,  2004, \apj, 614,
  679

\bibitem[\protect\citeauthoryear{{Maeder} \& {Meynet}}{{Maeder} \&
  {Meynet}}{1989}]{1989A&A...210..155M}
{Maeder} A.,  {Meynet} G.,  1989, \aap, 210, 155

\bibitem[\protect\citeauthoryear{{Matteucci} \& {Recchi}}{{Matteucci} \&
  {Recchi}}{2001}]{2001ApJ...558..351M}
{Matteucci} F.,  {Recchi} S.,  2001, \apj, 558, 351

\bibitem[\protect\citeauthoryear{{McIntosh}, {Zabludoff}, {Rix} \&
  {Caldwell}}{{McIntosh} et~al.}{2005}]{2005ApJ...619..193M}
{McIntosh} D.~H.,  {Zabludoff} A.~I.,  {Rix} H.-W.,    {Caldwell} N.,  2005,
  \apj, 619, 193

\bibitem[\protect\citeauthoryear{{McNamara}, {Rafferty}, {Birzan}, {Steiner},
  {Wise}, {Nulsen}, {Carilli}, {Ryan} \& {Sharma}}{{McNamara}
  et~al.}{2006}]{2006astro.ph..4044M}
{McNamara} B.~R.,  {Rafferty} D.~A.,  {Birzan} L.,  {Steiner} J.,  {Wise}
  M.~W.,  {Nulsen} P.~E.~J.,  {Carilli} C.~L.,  {Ryan} R.,    {Sharma} M.,
  2006, ArXiv Astrophysics e-prints

\bibitem[\protect\citeauthoryear{{Menci}, {Cavaliere}, {Fontana}, {Giallongo}
  \& {Poli}}{{Menci} et~al.}{2002}]{2002ApJ...575...18M}
{Menci} N.,  {Cavaliere} A.,  {Fontana} A.,  {Giallongo} E.,    {Poli} F.,
  2002, \apj, 575, 18

\bibitem[\protect\citeauthoryear{{Metzler} \& {Evrard}}{{Metzler} \&
  {Evrard}}{1994}]{1994ApJ...437..564M}
{Metzler} C.~A.,  {Evrard} A.~E.,  1994, \apj, 437, 564

\bibitem[\protect\citeauthoryear{{Moran}, {Ellis}, {Treu}, {Smail}, {Dressler},
  {Coil} \& {Smith}}{{Moran} et~al.}{2005}]{2005ApJ...634..977M}
{Moran} S.~M.,  {Ellis} R.~S.,  {Treu} T.,  {Smail} I.,  {Dressler} A.,  {Coil}
  A.~L.,    {Smith} G.~P.,  2005, \apj, 634, 977

\bibitem[\protect\citeauthoryear{{Nagai} \& {Kravtsov}}{{Nagai} \&
  {Kravtsov}}{2005}]{2005ApJ...618..557N}
{Nagai} D.,  {Kravtsov} A.~V.,  2005, \apj, 618, 557

\bibitem[\protect\citeauthoryear{Navarro, Frenk \& White}{Navarro
  et~al.}{1996}]{NA96.1}
Navarro J.,  Frenk C.,    White S.,  1996, ApJ, 462, 563

\bibitem[\protect\citeauthoryear{{Nomoto}, {Iwamoto}, {Nakasato}, {Thielemann},
  {Brachwitz}, {Tsujimoto}, {Kubo} \& {Kishimoto}}{{Nomoto}
  et~al.}{1997}]{1997NuPhA.621..467N}
{Nomoto} K.,  {Iwamoto} K.,  {Nakasato} N.,  {Thielemann} F.-K.,  {Brachwitz}
  F.,  {Tsujimoto} T.,  {Kubo} Y.,    {Kishimoto} N.,  1997, Nuclear Physics A,
  621, 467

\bibitem[\protect\citeauthoryear{{Oemler}}{{Oemler}}{1974}]{1974ApJ...194....1%
O}
{Oemler} A.~J.,  1974, \apj, 194, 1

\bibitem[\protect\citeauthoryear{{Pimbblet}, {Smail}, {Edge}, {O'Hely}, {Couch}
  \& {Zabludoff}}{{Pimbblet} et~al.}{2006}]{2006MNRAS.366..645P}
{Pimbblet} K.~A.,  {Smail} I.,  {Edge} A.~C.,  {O'Hely} E.,  {Couch} W.~J.,
  {Zabludoff} A.~I.,  2006, \mnras, 366, 645

\bibitem[\protect\citeauthoryear{{Poggianti}}{{Poggianti}}{2004}]{2004bdmh.con%
fE.104P}
{Poggianti} B.,  2004, in {Dettmar} R.,  {Klein} U.,   {Salucci} P.,  eds,
  Baryons in Dark Matter Halos {Evolution of galaxies in clusters}

\bibitem[\protect\citeauthoryear{{Popesso}, {Biviano}, {B{\"o}hringer} \&
  {Romaniello}}{{Popesso} et~al.}{2006a}]{2006A&A...445...29P}
{Popesso} P.,  {Biviano} A.,  {B{\"o}hringer} H.,    {Romaniello} M.,  2006a,
  \aap, 445, 29

\bibitem[\protect\citeauthoryear{{Popesso}, {Biviano}, {B{\"o}hringer} \&
  {Romaniello}}{{Popesso} et~al.}{2006b}]{2006astro.ph..6260P}
{Popesso} P.,  {Biviano} A.,  {B{\"o}hringer} H.,    {Romaniello} M.,  2006b,
  ArXiv Astrophysics e-prints

\bibitem[\protect\citeauthoryear{{Popesso}, {Biviano}, {B{\"o}hringer},
  {Romaniello} \& {Voges}}{{Popesso} et~al.}{2005}]{2005A&A...433..431P}
{Popesso} P.,  {Biviano} A.,  {B{\"o}hringer} H.,  {Romaniello} M.,    {Voges}
  W.,  2005, \aap, 433, 431

\bibitem[\protect\citeauthoryear{{Prugniel} \& {Simien}}{{Prugniel} \&
  {Simien}}{1996}]{1996A&A...309..749P}
{Prugniel} P.,  {Simien} F.,  1996, \aap, 309, 749

\bibitem[\protect\citeauthoryear{{Rafferty}, {McNamara}, {Nulsen} \&
  {Wise}}{{Rafferty} et~al.}{2006}]{2006astro.ph..5323R}
{Rafferty} D.~A.,  {McNamara} B.~R.,  {Nulsen} P.~E.~J.,    {Wise} M.~W.,
  2006, ArXiv Astrophysics e-prints

\bibitem[\protect\citeauthoryear{{Ramella}, {Boschin}, {Geller}, {Mahdavi} \&
  {Rines}}{{Ramella} et~al.}{2004}]{2004AJ....128.2022R}
{Ramella} M.,  {Boschin} W.,  {Geller} M.~J.,  {Mahdavi} A.,    {Rines} K.,
  2004, \aj, 128, 2022

\bibitem[\protect\citeauthoryear{{Recchi}, {Matteucci} \& {D'Ercole}}{{Recchi}
  et~al.}{2001}]{2001MNRAS.322..800R}
{Recchi} S.,  {Matteucci} F.,    {D'Ercole} A.,  2001, \mnras, 322, 800

\bibitem[\protect\citeauthoryear{{Renzini} \& {Voli}}{{Renzini} \&
  {Voli}}{1981}]{1981A&A....94..175R}
{Renzini} A.,  {Voli} M.,  1981, \aap, 94, 175

\bibitem[\protect\citeauthoryear{{Rines}, {Geller}, {Diaferio}, {Kurtz} \&
  {Jarrett}}{{Rines} et~al.}{2004}]{2004AJ....128.1078R}
{Rines} K.,  {Geller} M.~J.,  {Diaferio} A.,  {Kurtz} M.~J.,    {Jarrett}
  T.~H.,  2004, \aj, 128, 1078

\bibitem[\protect\citeauthoryear{{Romeo}, {Portinari} \&
  {Sommer-Larsen}}{{Romeo} et~al.}{2005}]{2005MNRAS.361..983R}
{Romeo} A.~D.,  {Portinari} L.,    {Sommer-Larsen} J.,  2005, \mnras, 361, 983

\bibitem[\protect\citeauthoryear{{Salpeter}}{{Salpeter}}{1955}]{1955ApJ...121.%
.161S}
{Salpeter} E.~E.,  1955, \apj, 121, 161

\bibitem[\protect\citeauthoryear{{Schechter}}{{Schechter}}{1976}]{1976ApJ...20%
3..297S}
{Schechter} P.,  1976, \apj, 203, 297

\bibitem[\protect\citeauthoryear{{Shandarin} \& {Zeldovich}}{{Shandarin} \&
  {Zeldovich}}{1989}]{1989RvMP...61..185S}
{Shandarin} S.~F.,  {Zeldovich} Y.~B.,  1989, Reviews of Modern Physics, 61,
  185

\bibitem[\protect\citeauthoryear{{Sijacki} \& {Springel}}{{Sijacki} \&
  {Springel}}{2006}]{2006MNRAS.366..397S}
{Sijacki} D.,  {Springel} V.,  2006, \mnras, 366, 397

\bibitem[\protect\citeauthoryear{{Somerville} \& {Primack}}{{Somerville} \&
  {Primack}}{1999}]{1999MNRAS.310.1087S}
{Somerville} R.~S.,  {Primack} J.~R.,  1999, \mnras, 310, 1087

\bibitem[\protect\citeauthoryear{{Springel}}{{Springel}}{2005}]{2005MNRAS.364.%
1105S}
{Springel} V.,  2005, \mnras, 364, 1105

\bibitem[\protect\citeauthoryear{{Springel} \& {Hernquist}}{{Springel} \&
  {Hernquist}}{2002}]{2002MNRAS.333..649S}
{Springel} V.,  {Hernquist} L.,  2002, \mnras, 333, 649

\bibitem[\protect\citeauthoryear{{Springel} \& {Hernquist}}{{Springel} \&
  {Hernquist}}{2003a}]{2003MNRAS.339..289S}
{Springel} V.,  {Hernquist} L.,  2003a, \mnras, 339, 289

\bibitem[\protect\citeauthoryear{{Springel} \& {Hernquist}}{{Springel} \&
  {Hernquist}}{2003b}]{2003MNRAS.339..312S}
{Springel} V.,  {Hernquist} L.,  2003b, \mnras, 339, 312

\bibitem[\protect\citeauthoryear{Springel, White, Tormen \& Kauffmann}{Springel
  et~al.}{2001}]{SP01.2}
Springel V.,  White S.,  Tormen G.,    Kauffmann G.,  2001, MNRAS, 328, 726

\bibitem[\protect\citeauthoryear{{Springel}, {White}, {Jenkins}, {Frenk},
  {Yoshida}, {Gao}, {Navarro}, {Thacker}, {Croton}, {Helly}, {Peacock}, {Cole},
  {Thomas}, {Couchman}, {Evrard}, {Colberg} \& {Pearce}}{{Springel}
  et~al.}{2005}]{2005Natur.435..629S}
{Springel} V.,  {White} S.~D.~M.,  {Jenkins} A.,  {Frenk} C.~S.,  {Yoshida} N.,
   {Gao} L.,  {Navarro} J.,  {Thacker} R.,  {Croton} D.,  {Helly} J.,
  {Peacock} J.~A.,  {Cole} S.,  {Thomas} P.,  {Couchman} H.,  {Evrard} A.,
  {Colberg} J.,    {Pearce} F.,  2005, \nat, 435, 629

\bibitem[\protect\citeauthoryear{Springel, Yoshida \& White}{Springel
  et~al.}{2001}]{SP01.1}
Springel V.,  Yoshida N.,    White S.,  2001, New Astronomy, 6, 79

\bibitem[\protect\citeauthoryear{{Stadel}}{{Stadel}}{2001}]{2001PhDT........21%
S}
{Stadel} J.~G.,  2001, Ph.D.~Thesis

\bibitem[\protect\citeauthoryear{{Stoughton}, {Lupton}, {Bernardi} \& {et
  al.}}{{Stoughton} et~al.}{2002}]{2002AJ....123..485S}
{Stoughton} C.,  {Lupton} R.~H.,  {Bernardi} M.,    {et al.} 2002, \aj, 123,
  485

\bibitem[\protect\citeauthoryear{{Strazzullo}, {Rosati}, {Stanford}, {Lidman},
  {Nonino}, {Demarco}, {Eisenhardt}, {Ettori}, {Mainieri} \&
  {Toft}}{{Strazzullo} et~al.}{2006}]{2006astro.ph..1165S}
{Strazzullo} V.,  {Rosati} P.,  {Stanford} S.~A.,  {Lidman} C.,  {Nonino} M.,
  {Demarco} R.,  {Eisenhardt} P.~E.,  {Ettori} S.,  {Mainieri} V.,    {Toft}
  S.,  2006, ArXiv Astrophysics e-prints

\bibitem[\protect\citeauthoryear{{Sutherland} \& {Dopita}}{{Sutherland} \&
  {Dopita}}{1993}]{1993ApJS...88..253S}
{Sutherland} R.~S.,  {Dopita} M.~A.,  1993, \apjs, 88, 253

\bibitem[\protect\citeauthoryear{{Terlevich}, {Caldwell} \&
  {Bower}}{{Terlevich} et~al.}{2001}]{2001MNRAS.326.1547T}
{Terlevich} A.~I.,  {Caldwell} N.,    {Bower} R.~G.,  2001, \mnras, 326, 1547

\bibitem[\protect\citeauthoryear{{Thielemann}, {Nomoto} \&
  {Hashimoto}}{{Thielemann} et~al.}{1996}]{1996ApJ...460..408T}
{Thielemann} F.-K.,  {Nomoto} K.,    {Hashimoto} M.-A.,  1996, \apj, 460, 408

\bibitem[\protect\citeauthoryear{{Thomas} \& {Couchman}}{{Thomas} \&
  {Couchman}}{1992}]{1992MNRAS.257...11T}
{Thomas} P.~A.,  {Couchman} H.~M.~P.,  1992, \mnras, 257, 11

\bibitem[\protect\citeauthoryear{{Tormen}, {Bouchet} \& {White}}{{Tormen}
  et~al.}{1997}]{1997MNRAS.286..865T}
{Tormen} G.,  {Bouchet} F.~R.,    {White} S.~D.~M.,  1997, \mnras, 286, 865

\bibitem[\protect\citeauthoryear{{Tornatore}, {Borgani}, {Matteucci}, {Recchi}
  \& {Tozzi}}{{Tornatore} et~al.}{2004}]{2004MNRAS.349L..19T}
{Tornatore} L.,  {Borgani} S.,  {Matteucci} F.,  {Recchi} S.,    {Tozzi} P.,
  2004, \mnras, 349, L19

\bibitem[\protect\citeauthoryear{{van Dokkum}, {Franx}, {Fabricant},
  {Illingworth} \& {Kelson}}{{van Dokkum} et~al.}{2000}]{2000ApJ...541...95V}
{van Dokkum} P.~G.,  {Franx} M.,  {Fabricant} D.,  {Illingworth} G.~D.,
  {Kelson} D.~D.,  2000, \apj, 541, 95

\bibitem[\protect\citeauthoryear{Voges}{Voges}{1992}]{VO92.1}
Voges W.,  1992, in Environment observation and climate modelling through
  international space projects. Space sciences with particular emphasis on
  high-energy astrophysics The rosat all-sky x-ray survey.
ESA, p.~9

\bibitem[\protect\citeauthoryear{{Wake}, {Collins}, {Nichol}, {Jones} \&
  {Burke}}{{Wake} et~al.}{2005}]{2005ApJ...627..186W}
{Wake} D.~A.,  {Collins} C.~A.,  {Nichol} R.~C.,  {Jones} L.~R.,    {Burke}
  D.~J.,  2005, \apj, 627, 186

\bibitem[\protect\citeauthoryear{White}{White}{1996}]{WH96.1b}
White S.,  1996, in Schaeffer R.,  Silk J.,  Spiro M.,   Zinn-Justin J.,  eds,
  Cosmology and Large-Scale Structure The formation and evolution of galaxies.
Elsevier, Dordrecht, p.~395

\bibitem[\protect\citeauthoryear{{Woosley} \& {Weaver}}{{Woosley} \&
  {Weaver}}{1995}]{1995ApJS..101..181W}
{Woosley} S.~E.,  {Weaver} T.~A.,  1995, \apjs, 101, 181

\bibitem[\protect\citeauthoryear{{Yoshida}, {Sheth} \& {Diaferio}}{{Yoshida}
  et~al.}{2001}]{2001MNRAS.328..669Y}
{Yoshida} N.,  {Sheth} R.~K.,    {Diaferio} A.,  2001, \mnras, 328, 669

\bibitem[\protect\citeauthoryear{Yoshida, Stoehr, Springel \& White}{Yoshida
  et~al.}{2002}]{YO02.1}
Yoshida N.,  Stoehr F.,  Springel V.,    White S.,  2002, MNRAS, 334, 762

\bibitem[\protect\citeauthoryear{{Zanni}, {Murante}, {Bodo}, {Massaglia},
  {Rossi} \& {Ferrari}}{{Zanni} et~al.}{2005}]{2005A&A...429..399Z}
{Zanni} C.,  {Murante} G.,  {Bodo} G.,  {Massaglia} S.,  {Rossi} P.,
  {Ferrari} A.,  2005, \aap, 429, 399

\end{thebibliography}

\section*{Appendix. Testing numerical effects on the galaxy stellar
  mass function} In this Appendix we discuss the stability, against
possible numerical effects, of the stellar mass function of the
galaxies identified inside simulated clusters. In particular, we will
focus the analysis on {\em (a)} the effect of changing the softening
of the gravitational force, to control the possible presence of
spurious numerical heating \citep[e.g.,][]{1992MNRAS.257...11T}; {\em
(b)} the effect of mass and force resolution. As for the softening
choice, it is known that using too small values may induce spurious
heating of the gas particles by two--body collisions, thereby
inhibiting gas cooling inside small halos. On the other hand,
increasing it to too large a value also reduces the number of galaxies
as a consequence of the lower number of resolved small halos
\citep{2006MNRAS.367.1641B}. As for the resolution, increasing it has
the effect to better resolve the low end of the mass function and, in
general, is expected to produce a more reliable galaxy population.

The tests described in this Appendix are based on three cluster sized
halos, which have been resimulated by varying mass and force
resolution. These clusters have virial masses in the range
$(1.6$--$2.9)\times 10^{14}\msun$ and are described in detail by
\cite{2006MNRAS.367.1641B}. They have been simulated for the same
cosmological model of the clusters described in this paper, but with a
lower normalization of the power spectrum, $\sigma_8=0.8$. At the
lowest resolution, the mass of the gas particle is $m_{\rm gas}\simeq
6.9\times 10^8\msun$ and the Plummer--equivalent softening for
gravitational forces is set to $\epsilon=7.5h^{-1}$kpc at $z=0$.  
We point out that the simulations analysed in this paper and described
in Section \ref{s:simul} have a mass resolution which is better than
the above one by about a factor of four. Runs at increasingly higher
resolution have been performed by decreasing the particle masses by a
factor 3, 10 and 45, with the softening correspondingly decreased
according to the $m^{1/3}$ scaling. Therefore, at the highest
resolution, it is $m_{\rm gas}\simeq 1.5\times 10^7\msun$ and
$\epsilon=2.1h^{-1}$kpc. This mass resolution is $\simeq 11$ times
higher than used for the set of simulations analysed in this
paper. For the most massive of these three clusters, we have also
repeated the simulation at 10 times the basic mass resolution with
four different choices of the gravitational softening. In particular,
we have decreased it by a factor two, with respect to the standard
choice, and increased it by a factor two and four. As such, this set
of simulations allows us to verify the stability of the galaxy stellar
mass function against changing mass resolution and the choice for the
gravitational softening.

The simulations have been performed with the original prescription for
star formation and feedback presented by \cite{2003MNRAS.339..289S},
with a wind speed $v_w\simeq 480\vel$, therefore comparable to that
assumed for the standard feedback in this paper. However, those runs
did not include the prescription for stellar evolution and chemical
enrichment, which we used here. Since each gas particle is allowed to
spawn two star particles, the latter have a mass which is half of that
of the parent gas particle. As discussed by
\cite{2006MNRAS.367.1641B}, this series of runs produces an amount of
stars within the virial radius of the clusters, which is almost
independent of the resolution, thereby preventing the runaway of
cooling with increasing resolution.

\begin{figure}
\psfig{file=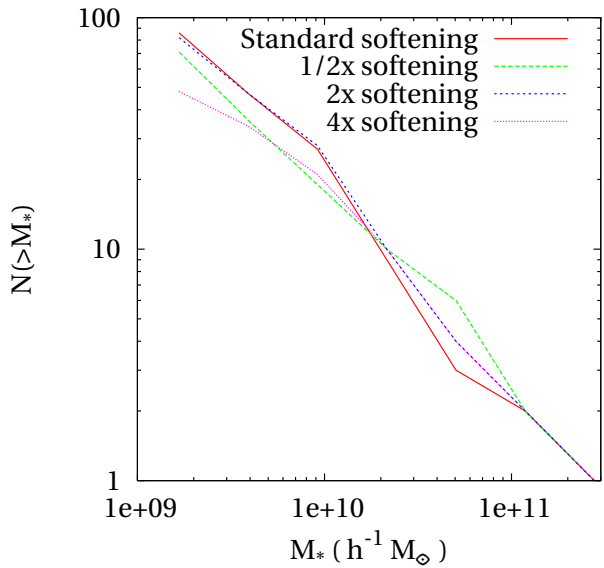,angle=-90,width=11cm} 
\caption{The cumulative stellar mass function of the galaxies
  identified within the virial radii of the most massive among the
  three simulated clusters described in the Appendix (see also
  \protect\citealt{2006MNRAS.367.1641B}). All the simulations have been done
  at fixed mass resolution, which correspond to an increase by a
  factor 10 with respect to the basic resolution (i.e., $m_{\rm
    gas}\simeq 6.9\times 10^7\msun$; see text). The four curves
  correspond to the different choices for the gravitational softening.
  The labels indicate the factor by which the softening has been
  changed, with respect to the standard choice of $3.5h^{-1}$kpc.}
\label{fi:mfsoft}
\end{figure}

We show in Figure \ref{fi:mfsoft} the effect of varying the softening
on the cumulative stellar mass function of the galaxies identified
inside the virial radius. As expected, decreasing the softening to
half the standard value has the effect of suppressing the low end of
the mass function, $M\mincir 2\times 10^{10}h^{-1}M_\odot$, as a
consequence of spurious numerical heating of gas. On the other hand,
increasing the softening by a factor four also induces a suppression
of low--mass galaxies, as a consequence of the lack of resolution. At
larger masses, using a too small softening has the effect of
increasing the mass function, although the rather small number of
galaxies in the high mass end prevents from detecting systematic
trends. These results demonstrate that our lack of galaxies can not be
explained by a non--optimal choice of the gravitational force
softening.

\begin{figure}
\psfig{file=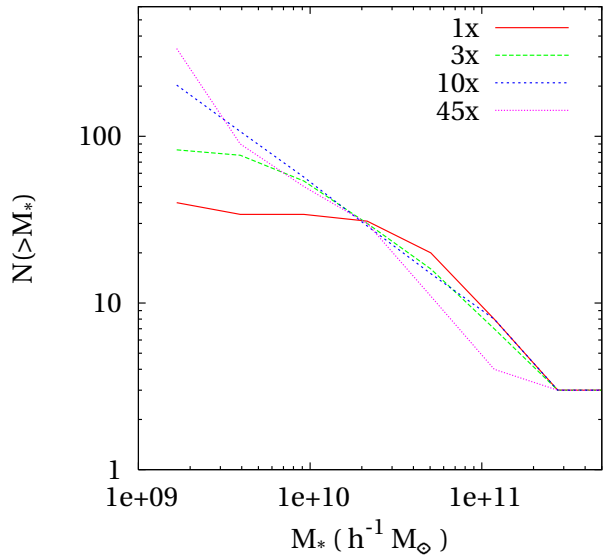,angle=-90,width=11cm} 
\caption{The combined cumulative stellar mass function of the galaxies
  identified within the virial radii of the three clusters. The four
  curves correspond to the different resolutions at which the clusters
  have been simulated. Continuous, long--dashed, short--dashed and
  dotted curves are for the simulations at progressively increasing
  resolution. The labels indicate the factor by which mass resolution
  is increased, with respect to the lowest resolution run (1x).}
\label{fi:mftark} 
\end{figure}

As for the effect of resolution, we plot in Figure \ref{fi:mftark} the
combined cumulative stellar mass function for all the galaxies
identified within the virial radii of the three clusters, simulated at
four different resolutions. 
The first apparent effect of increasing resolution is that of
steepening the mass function in the low mass end. In the mass
range where galaxies are identified with at least 64 star particles at
the different resolutions, which correspond to $M_*\simeq
2.2\times 10^{10}h^{-1}M_\odot$ for the lowest resolution run, the
mass functions have a weaker dependence on resolution, with a
decreasing trend of the high end of the mass function. This steepening
of the high end of the mass function at increasing resolution is
the consequence of the reduction of overmerging, which makes small
halos surviving more efficiently and, therefore, prevents their
disruption and accretion inside massive halos. This result
demonstrates that, at least at the highest resolution reached in this
test, also resolution is not the reason for the too low number of
galaxies found in the simulated clusters, when compared to
observations (see discussion in Sect. \ref{s:ml}).

\end{document}